\documentclass[traditabstract]{aa} % for the abstract without structuration 
\usepackage{natbib}
\usepackage{graphicx}
\usepackage{txfonts}
\begin{document}

    \title{Distribution of refractory and volatile elements in CoRoT exoplanet host stars
\thanks{Based on observations made with the 2.2m telescope 
at the European Southern Observatory (La Silla, Chile).}}     
    
      \author{C. Chavero\inst{1},  R. de la Reza\inst{1}, R.C. Domingos\inst{3}, 
N.A. Drake\inst{2}, C.B. Pereira\inst{1} \&  O. C. Winter\inst{3}  }

     \offprints{C. Chavero \email{carolina@on.br}}

     \institute{ Observat\'orio Nacional, Rua Jos\'e Cristino 77, S\~ao
Cristov\~ao,  20921-400, Rio de Janeiro, Brazil \\ emails: carolina@on.br, delareza@on.br,
 drake@on.br, claudio@on.br
      \and
Sobolev Astronomical Institute, St.~Petersburg State University, 
Universitetski pr.~28, St.~Petersburg 198504, Russia
\and   Univ Estadual Paulista - UNESP, Grupo de Din\^amica Orbital \& Planetologia, Guaratinguet\'a, CEP 12.516-410, Brazil
  	 \\ emails: rcassia@feg.unesp.br, ocwinter@pq.cnpq.br}

     \date{Received ; accepted }

\abstract
{
The relative distribution of  abundances of refractory, intermediate, and volatile
elements in stars with planets can be an important tool for investigating the internal
migration of a giant planet. This migration can lead to the  accretion of planetesimals
and the selective  enrichment of the star with these elements. 
We report on a spectroscopic determination of the atmospheric parameters
and chemical abundances of the parent stars in transiting
planets CoRoT-2b and CoRoT-4b. Adding data for CoRoT-3 and CoRoT-5 from the
literature, we find a  flat distribution of  the relative abundances as a function of their
condensation temperatures. 
For CoRoT-2, the   relatively  high lithium abundance and intensity of  
its Li\,{\sc i} resonance line permit us to propose an age of 120 Myr, 
making this stars one of the youngest stars  with planets to date.
We introduce a new methodology to investigate a relation between the  abundances of these stars 
and the internal migration of their planets. By simulating  the internal
migration of a planet in a disk formed only by planetesimals, we are able, for the first
time, to separate the stellar fractions of refractory, intermediate, and volatile rich
planetesimals accreting onto the central star. Intermediate and volatile element
fractions enriching the star are similar and much larger than those of pure refractory
ones. This result is opposite to what has been considered in the literature for the
accreting self-enrichment processes of stars with planets. We also show that these results
are highly dependent on the model adopted for the disk distribution regions in terms of refractory,
intermediate, and also volatile elements and other parameters considered.
We note however, that this self-enrichment mechanism
is only efficient during the first 20 -- 30 Myr or later in the
lifetime of the disk when the surface convection layers of the
central star  for the first time attain  its minimum size configuration. 
}

\authorrunning{Chavero et al.}
\titlerunning{Abundance distributions in CoRoT exoplanet host stars
}

\maketitle

%________________________________________________________________
\section{Introduction}

\cite{gonzalez97} and \cite{santos01} first identified a metallicity excess
in stars with giant planets (SWP) relative  to stars without planets. The tendency has been
confirmed using  large samples of stars. Nevertheless, planets have also been observed around
low metallicity main-sequence stars \citep{cochran07}. We note, however, that such a large
statistical metal excess found among dwarf stars is not present in  giant stars with
planets \citep{pasquini07}.

 Using iron as a primary reference element, this metal excess, has been the source of a 
number of studies trying to explain this property. Two scenarios have been invoked: a 
self-enrichment mechanism of normal metal stars and a primordial scenario, in which SWP 
are the result of the formation of entire metal-rich stars in equally metal-rich natal clouds. 
Which scenario is  correct  remains an open question to this day.

What about  CoRoT exoplanet host stars? 
Although there are few cases with which  to perform a statistical study, these stars exhibit
a different distributions: the majority  have  solar abundances, one has a mild metal excess, 
and one  has a  low metal abundance. Even at this early stage,  we are tempted to
speculate that CoRoT is exploring a different backyard of the Galaxy with different 
properties from those of the Solar vicinity. Where  does the bulk of the metal excess of 
the SWP mentioned above  originate?

The relative distribution of the stellar abundances of refractory
elements, with a high condensation temperature ($T_{\rm C}$), with respect to the
volatile elements (low $T_{\rm C}$), has been widely used in the literature as
a tool to investigate the nature of  metal enrichment in SWP
(see review papers of \cite{gonzalez03,gonzalez06b}). 

This enrichment mechanism should be triggered  by the action of a hot Jupiter.
The observed pile up of hot Jupiters is believed to be the result of migration. 
These planets formed further away in the protoplanetary nebula and migrated afterwards 
to the small orbital distances at which they are observed. During inward migration, 
material (planetesimals) from the disk, depleted in H and He, could be accreted onto the star.
The presence of a positive slope of [X/H] versus $T_{\rm C}$, in which the  abundances of volatile 
elements  are lower than that of the refractory abundances, has
been considered as a possible signature of a self-enrichment mechanism \citep{smith01}.
This is based on  the never proven assumption that stellar accretion only favors near star
disk regions rich in refractory elements.

In \cite{winter07}, we explored by means of numerical simulations  stellar accretion 
caused by an inward forced planetary migration in a disk formed only by planetesimals
that cause  a metallic self-enrichment of the central star photosphere. At present, we
attempt to identify this stellar accretion by determining the origin in the disk of 
these particles (planetesimals). In this way, we can separate the contributions
of rocky material enriched differentially with refractory, intermediate, and volatile elements. 
In this work, we apply this methodology to three different systems, namely: 
CoRoT-2 \citep{alonso08}, 3 \citep{deleuil08}, and 4 \citep{aigrain08}.
The main purpose of this work is to analyze the mentioned slopes of [Fe/H] vs. $T_{\rm C}$ 
corresponding to CoRoT-2, CoRoT-3, CoRoT-4, and CoRoT-5 and their interpretations. 
For cases 2 and 4, we obtained new high resolution spectra. Cases 3 and 5 are taken
from the  literature.

\section{Observations}

The high-resolution spectra of CoRoT-2 and CoRoT-4  analyzed in this work were obtained
with the FEROS (Fiberfed Extended Range Optical Spectrograph) echelle spectrograph 
\citep{Kaufer99} of the 2.2 m ESO telescope at La Silla (Chile), on October 23th, 2008.
The FEROS spectral resolving power is R$=$48 000, corresponding to 2.2 pixel of $15~\mu$m,
and the wavelength coverage ranges from 3500  to 9200 ~\AA. 
The nominal  signal-to-noise ratio (S/N) was evaluated by measuring the rms-flux fluctuation
in selected continuum windows, and the typical values were S/N 55-60 after  $2\times 3600$ s of 
integration time. The spectra were reduced with the MIDAS pipeline reduction 
package consisting of the following standard steps: CCD bias correction, 
flat-fielding, spectrum extraction, wavelength calibration, 
correction of barycentric velocity, and spectrum rectification.

\section{Abundance analysis of CoRoT-2 and CoRoT-4}

\subsection{Stellar parameters and iron abundances}

Determination of the basic, effective temperature ($T_{\rm eff}$), surface gravity (log {\rm g}), microturbulence ($\xi$$_t$),
metallicity ([Fe/H]), and  the  abundance analysis were derived by means of the  standard approach of the local thermodynamic equilibrium (LTE) using
a revised version (2002) of the code MOOG \citep{Sneden73} and a grid of \cite{Kurucz93} ATLAS9 atmospheres,  which include overshooting. 

The atmospheric parameters were obtained from the iron lines (Fe\,{\sc i} and F\,{\sc ii} )  by iterating until the correlation coefficients
between $\log\varepsilon$(Fe\,{\sc i}) and lower excitation potential ($\chi l$), and between  $\log\varepsilon$(Fe\,{\sc i}) and 
reduced equivalent width ($\log ({W_\lambda}/\lambda$) were zero, and the mean abundance given by Fe\,{\sc i} and Fe\,{\sc ii} lines were similar.
The iron lines taken from \cite{lambert96} and \cite{santos04} were carefully chosen  
by verifying that each line was not too strong, and checked for possible blending.

 We  adopted new $\log{gf}$ values for the iron lines. These values were computed from an inverted solar analysis using solar 
equivalent widths (EWs) measured from a solar spectrum taken with FEROS and a Kurucz grid model for 
the Sun (Kurucz 1993) having ($T_{{\rm eff}}$, $\log{g}$, $\xi_{{\rm t}}$, $\log\varepsilon(\rm Fe)$)
 = (5777~K, 4.44 dex, 1.00 kms$^{-1}$, 7.47 dex). Table \ref{t:fe} contains the  linelist used, 
the  adopted parameters, and also includes the EWs measured in each star. 
Table  \ref{t:parameters} shows the atmospheric parameters derived by us and those found by other authors. 

We determined equivalent widths using the ``splot'' task of IRAF \footnote{IRAF is distributed by the National 
Optical Astronomy Observatory, which is operated by the Association of Universities for Research in Astronomy, Inc.,
under cooperative agreement with the National Science Foundation.}. Abundances were computed using the ABFIND driver in MOOG.

\begin{table}[!]
\caption{Atomic parameters of the iron lines }

\label{t:fe}
\begin{tabular}{l c c  c | c|  c        }
\hline\hline
Species& $\lambda$ & $\chi$$_l$ & log $\it {gf}$ & EW (m\AA) & EW (m\AA)    \\
       &           &            &            & CoRoT-2     & CoRoT-4        \\
\hline

Fe\,{\sc i}&5044.22  &  2.85  &   -2.043    &   88.0   &  66.7   \\
&5198.72             &  2.22  &   -2.176    &          &  94.5   \\
&5242.49             &  3.63  &   -1.130    &  102.3   &  84.1   \\
&5322.05             &  2.28  &   -2.900    &   77.2   &  56.3   \\
&5638.27             &  4.22  &   -0.809    &   94.1   &  77.6   \\
&5731.77             &  4.26  &   -1.124    &   73.0   &  53.2   \\
&5806.73             &  4.61  &   -0.893    &   70.3   &  51.9   \\
&5852.22             &  4.55  &   -1.185    &   52.2   &  36.2   \\
&5855.08             &  4.61  &   -1.530    &   26.1   &         \\
&5856.09             &  4.29  &   -1.568    &   39.3   &         \\
&6027.06             &  4.08  &   -1.190    &   80.0   &  61.6   \\
&6056.01             &  4.73  &   -0.500    &   92.2   &  73.2   \\
&6079.01             &  4.65  &   -1.010    &          &  39.7   \\
&6089.57             &  5.02  &   -0.884    &   45.1   &  36.5   \\
&6151.62             &  2.18  &   -3.300    &   70.6   &  40.2   \\
&6157.73             &  4.07  &   -1.240    &   76.5   &  61.3   \\
&6159.38             &  4.61  &   -1.860    &          &         \\
&6165.36             &  4.14  &   -1.500    &   53.3   &  42.0   \\
&6180.21             &  2.73  &   -2.635    &   71.7   &  54.0   \\
&6188.00             &  3.94  &   -1.635    &   58.7   &  45.6   \\
&6200.32             &  2.61  &   -2.400    &   98.0   &  65.9   \\
&6213.44             &  2.22  &   -2.592    &  102.0   &  76.2   \\
&6226.74             &  3.88  &   -2.070    &   33.6   &         \\
&6229.24             &  2.84  &   -2.880    &   45.5   &         \\
&6265.14             &  2.18  &   -2.574    &  111.0   &  77.0   \\
&6270.23             &  2.86  &   -2.588    &   68.5   &  49.0   \\
&6380.75             &  4.19  &   -1.320    &   66.9   &  51.9   \\
&6392.54             &  2.28  &   -3.933    &   25.1   &         \\
&6498.94             &  0.96  &   -4.632    &   63.5   &  38.0   \\
&6627.55             &  4.55  &   -1.480    &   32.1   &         \\
&6703.57             &  2.76  &   -3.027    &   43.1   &  31.3   \\
&6726.67             &  4.61  &   -1.050    &   51.4   &         \\
&6750.16             &  2.42  &   -2.621    &   93.2   &  75.4   \\
&6752.71             &  4.64  &   -1.230    &   47.0   &  36.7   \\
&6786.86             &  4.19  &   -1.900    &   30.0   &  21.0   \\
Fe\,{\sc ii}&5234.63 &  3.22  &   -2.238    &   96.4   & 100.8   \\
&  5991.38           &  3.15  &   -3.529    &          &  47.9   \\
&  6084.11           &  3.20  &   -3.767    &          &  33.3   \\
&  6149.25           &  3.89  &   -2.726    &   44.9   &  53.7   \\
&  6247.56           &  3.89  &   -2.342    &   64.5   &  72.2   \\
&  6369.46           &  2.89  &   -4.134    &   23.7   &         \\
&  6416.93           &  3.89  &   -2.640    &   43.1   &  57.0   \\
&  6432.69           &  2.89  &   -3.564    &   44.9   &  52.9   \\
&  7711.73           &  3.90  &   -2.538    &   49.8   &         \\
\hline                                                                                                                                                   
\end{tabular}                                                                                                                                      
\tablefoot{ Col 1. element; Col. 2: wavelength (in \AA\ );
 Col. 3: excitation energy of the lower energy level in the transition (in eV); Col. 4: oscillator strengths based
 on an inverse solar analysis.; Col. 5: EW for CoRoT-2; Col. 6: EW for CoRoT-4.
}
\end{table}

\begin{table*}[!]

\caption{Stellar parameters of the CoRoT exoplanets parent stars }
\label{t:parameters}
\begin{tabular}{lccccccc}
\hline
\hline
Star Name & $T_{\rm eff}  [K]$  & log g  &  $\xi$ $_t$ [km/s] & [Fe/H] &  [M/H] &  {\it v sin i} [km/s]  & References   \\
\hline
\\
\object{CoRoT-2 }  & 5696 $\pm$ 70 & 4.42 $\pm$ 0.12 & 1.71 $\pm$ 0.10  &  0.03  $\pm$ 0.06    & 0.03 $\pm$ 0.06 & 8.5 $\pm$ 1 & This work   \\
\object{CoRoT-4 }  & 6115 $\pm$ 70 & 4.30 $\pm$ 0.12 & 1.37  $\pm$ 0.10 &   0.12  $\pm$  0.05  & 0.10 $\pm$ 0.05 &  5.5 $\pm$ 1  & This work    \\
\hline
\\
CoRoT-2           & 5625 $\pm$ 120 & 4.30 $\pm$ 0.20 &  --             & --               & 0.0 $\pm$ 0.1 & 11.85 $\pm$ 0.45  &   \cite{bouchy08} \\
                  & 5608 $\pm$ 37  & 4.71 $\pm$ 0.20 & 1.49 $\pm$ 0.06 & 0.07 $\pm$ 0.04  &    --           &     --              &   \cite{ammlervon09}\\
CoRoT-4           & 6190 $\pm$ 60 & 4.41 $\pm$ 0.05 & 0.94  $\pm$ 0.05 &   --  & 0.05 $\pm$ 0.07 & 6.4 $\pm$ 1 & \cite{moutou08}     \\
\object{CoRoT-3 } & 6740 $\pm$ 140  & 4.22 $\pm$ 0.07   & -- & 0.03  $\pm$ 0.06  & -0.02 $\pm$ 0.06 & 17 $\pm$1 &\cite{deleuil08}   \\
\object{CoRoT-5 } & 6100 $\pm$  65  & 4.189 $\pm$0.03  & 0.91 $\pm$0.09 &  -0.25 $\pm$ 0.06 & -0.25 $\pm$ 0.06 & 1 $\pm$ 1 & \cite{rauer09}\\
\hline
\end{tabular}
\tablefoot{ Column 1: star name; Col. 2: effective temperature ; Col. 3: surface gravity
; Col. 4: microturbulence; Col. 5: metallicity; Col. 6: mean abundance (using element with  more than three lines);  Col. 7:
projected rotational velocity
Col. 8: references.}

\end{table*}

\subsection{Stellar abundances }

We present here results for 15  elements (O, Li, Na, Mg, Al, Si, S, Ca, Sc, Ti, Cr, Mn, Ni, Zn, and Ba),
plus Fe. The abundances of Na, Mg, Al, Si, Ca, Sc, Ti, Cr, Mn, Ni, Zn, and Ba  were derived from the analysis of 
EWs for several unblended lines, measured by Gaussian fitting. To derive semi-empirical atomic $\log{gf}$ values
for the lines of these elements, we used the EWs measured in the solar spectrum and performed an inverted solar analysis,
 as that carried out fort he iron element. 
The solar abundances of each element were taken from \cite{anders89}. Table  \ref{t:elements-lines} summarizes
the  linelist, the atomic parameters adopted, and the EWs measured in each star.
 For  Li, O, and S,  we used a spectral synthesis procedure. 

It is important to study the highly volatile CNO elements  because due to their low $T_{\rm C}$ values, they are indicators
of the cool disk zone below 371~K \citep{lodders03}. We took the condensation temperatures ($T_{\rm C}$)   
from \cite{lodders03}, covering a range between 70 and 1800~K.
An  attempt to measure abundances for an important volatile element, such as C, was unsuccessful.
A reliable calculation of the C abundance requires a higher  S/N than
used in this work. For instance, the C\,{\sc i} lines in the region of 7115 $\AA$  are hardly detectable
and are severely blended. Similar problems are found for other C\,{\sc i} lines, 
but we instead measured other volatile elements as S and  O for both stars.
The abundance ratios [X/H] (and [Fe/H]) for each element, averaged over all useful lines, are presented in Table
\ref{t:abundances}, along with the  number of lines used in each case.

\begin{table}[!]

\caption{Atomic parameters of the spectral lines used for each element. }
\label{t:elements-lines}
\begin{tabular}{l c c  c c  c  }
\hline\hline
Species                                               & $\lambda$ & $\chi$$_l$& log $\it {gf}$ & EW (m\AA)   &  EW (m\AA) \\
                                                      &           &           &           & CoRoT-2 & CoRoT-4 \\
\hline

Na\,{\sc i} &   5688.22  &  2.10  &  -0.678  &  133.7  &  110.3	\\
                    &   6154.23  &  2.10  &  -1.602  &   43.0  &   29.6	\\
                    &   6160.75  &  2.10  &  -1.316  &   64.0  &   43.0	\\
Mg\,{\sc i} &   5711.09  &  4.48  &  -1.706  &  120.0  &  96.0	\\
            &   6318.72  &  5.21  &  -1.996  &         &  36.0	\\
Al\,{\sc i} &   7835.31  &  4.02  &  -0.738  &   43.0  &  30.0	\\
            &   7836.13  &  4.02  &  -0.569  &   55.0  &  50.0	\\
Si\,{\sc i} &   5665.56  &  4.92  &  -1.980  &   50.1  &   33.0	\\
                    &   5948.55  &  5.08  &  -1.197  &         &   81.8	\\
                    &   6142.49  &  5.62  &  -1.518  &   35.3  &   28.0	\\
                    &   6145.02  &  5.62  &  -1.420  &   41.0  &  		\\
                    &   6721.84  &  5.86  &  -1.152  &   52.8  &   48.0	\\
Ca\,{\sc i} &   5581.97  &  2.52  &  -0.728  &  122.8  &   92.0	\\
                    &   5590.12  &  2.52  &  -0.815  &         &   82.0	\\
                    &   5867.56  &  2.93  &  -1.599  &    33.0 &  		\\
                    &   6166.44  &  2.52  &  -1.171  &    84.0 &   63.0	\\
                    &   6169.05  &  2.52  &  -0.812  &   113.6 &   90.0	\\
                    &   6169.56  &  2.52  &  -0.567  &   145.7 &  103.0	\\
                    &   6455.60  &  2.52  &  -1.420  &    72.1 &   45.0	\\
                    &   6471.66  &  2.53  &  -0.848  &   118.0 &   86.0	\\
Sc\,{\sc ii}&   5526.82  &  1.77  &   0.111  &   85.0  &    83.0	\\
            &   6245.65  &  1.51  &  -1.051  &   37.0  &    33.0	\\
            &   6604.60  &  1.36  &  -1.146  &         &    41.5	\\
Ti\,{\sc i}  &   5219.70  &  0.02  &  -2.268  &   46.4  &  		\\
                    &   5299.99  &  1.05  &  -1.425  &   29.4  &   16.0	\\
                    &   5866.46  &  1.07  &  -0.827  &   70.0  &   34.0	\\
                    &   6126.22  &  1.07  &  -1.398  &   32.7  &   12.0	\\
                    &   6258.11  &  1.44  &  -0.450  &   59.9  &  		\\
Cr\,{\sc i} &   5296.70  &  0.98  &  -1.411  &   115.0 &     84.0	\\
                    &   5300.75  &  0.98  &  -2.149  &    70.0 &     51.0	\\
                    &   5345.81  &  1.00  &  -1.036  &         &    106.5	\\
                    &   5348.33  &  1.00  &  -1.301  &         &     88.0	\\
                    &   5783.87  &  3.32  &  -0.220  &    60.0 &     40.0	\\
                    &   5787.93  &  3.32  &  -0.178  &    55.0 &     43.0	\\
Mn\,{\sc i} &   4265.92  &  2.94  &  -0.442  &         &  52.6	\\
                    &   4470.13  &  2.94  &  -0.585  &         &  43.0	\\
                    &   5399.47  &  3.85  &  -0.098  &    44.2 &  		\\
                    &   5432.54  &  0.00  &  -3.636  &    60.0 &  33.0	\\
Ni\,{\sc i} &   5084.11  &  3.68  &  -0.169  &   106.0 &   86.8	\\
                                                      &   6327.60  &  1.68  &  -3.100  &    43.2 &   38.1	\\
                                                      &   6643.64  &  1.68  &  -2.020  &   111.0 &   92.3	\\
                                                      &   6767.77  &  1.83  &  -2.170  &    91.0 &   77.5	\\
                                                      &   6772.32  &  3.66  &  -0.959  &    54.8 &   49.4	\\
                                                      &   7788.93  &  1.95  &  -1.951  &   101.2 &   88.3	\\
Zn\,{\sc i}			                      &   4722.16  &  4.03  &  -0.390  &    68.0 &  70.0	\\
                                                      &   4810.53  &  4.08  &  -0.310  &    87.0 &  75.0	\\
Ba\,{\sc ii}                                          &   5853.69  &  0.60  &  -0.848  &    90.5 &   75.0	\\
                                                      &   6141.73  &  0.70  &   0.146  &   160.0 &  130.0	\\
                                                      &   6496.90  &  0.60  &  -0.222  &   136.0 &  109.0	\\

\end{tabular}          
\tablefoot{Column 1: element; Col. 2: wavelength (in \AA\ );
 Col. 3: excitation energy of the lower energy level in the transition (in eV); Col. 4: oscillator strengths based on an inverse solar analysis.; Col. 5: EW for 
CoRoT-2; Col. 6: EW for CoRoT-4}
                                                                                        
\end{table}

\subsubsection{Abundance uncertainties}
 
The internal errors in the adopted effective temperature  and microturbulence  can be
determined from the uncertainty in the slope of both the Fe\,{\sc i} abundance versus excitation potential,
and the Fe\,{\sc i} abundance versus reduced equivalent width relations. 
The standard deviation in $\log g$ was inferred by changing this parameter around the adopted solution
until the  and   Fe\,{\sc ii} mean abundances differ by exactly one standard deviation of the  mean value 
of the  Fe\,{\sc i} abundance. These quantities are given in Table \ref{t:parameters}.

Errors and uncertainties  in the atmospheric parameters and determination for the continuum
can affect abundance measurements in various ways. We  tested the dependence of our results on 
atmospheric parameters obtained for CoRoT-2 and CoRoT4. 
\par Tables  \ref{t:errorcorot2} and  \ref{t:errorcorot4} show the  variation in the
abundance caused by $\Delta$ $T_{\rm eff}$, $\Delta$ log {\rm g}, $\Delta$ $\xi$$_t$, and $\Delta$ W.
The sixth column indicates the combined rms uncertainty of the second to fifth columns.
The last column indicates the observed abundance dispersion for  elements
whose abundances were derived using  three or more lines.

We can observe  that neutral elements are rather sensitive to temperature variations
 while single ionized elements are sensitive to the
variations in $\log g$. For the elements whose abundance is based
on stronger lines, such as barium, the error introduced by the microturbulence
is significant. For the elements analyzed by means of spectrum synthesis the same
technique was used, varying $T_{\rm eff}$, $\log g$, and $\xi$, then
independently computing the abundance changes introduced by the variation in
the above atmospheric parameters. 

\par The abundance uncertainties caused by the errors in the equivalent width 
measurements were computed from an expression provided by \cite{cayrel88}. The
errors in the equivalent widths are determined, essentially, by the S/N and the spectral resolution.
 In our case, having $R\approx 48\,000$ and
a typical $S/N$  of 55--60, the expected uncertainties in the equivalent
widths are about 3 m\AA. These error estimates were applied to the
measured $W_{\lambda}^{'}$s and the corresponding changes in the element
abundances are listed in Col. 5 of  Tables \ref{t:errorcorot2} and  \ref{t:errorcorot4}.

 The projected rotational velocities for both stars were determined by applying the synthetic spectra method 
to several single  and clean lines. The results presented in Table \ref{t:parameters} are the
following: for  CoRoT-2,  we obtain v$\sin i = 8.5 \pm 1$~km\,s$^{-1}$ and for CoRoT-4, we obtain 
v$\sin i = 5.5 \pm 1$~km\,s$^{-1}$.  These values can be compared with those obtained for the same 
stars  by other authors. For CoRoT-4,  \cite{moutou08} found v$ \sin i = 6.4 \pm 1.0$~km\,s$^{-1}$, which is
similar to our value. The case of CoRoT-2 is different. \cite{bouchy08} determined three independent values 
for this parameter from different procedures: 1) v$\sin i = 9.5 \pm 1.0$~km\,s$^{-1}$ from SOPHIE
cross-correlation functions (CCFs); 2) v$\sin i = 10.7 \pm 0.5$~km\,s$^{-1}$  from HARPS CCFs; and 3)
v$\sin i = 11.85 \pm 0.5$~km\,s$^{-1}$  by modeling the radial velocity anomaly occurring during the 
transit caused by the Rossiter-McLaughlin (RM) effect. The value obtained using SOPHIE 
are in closer agreement with our value since the difference is within the error bars.

Column 4 of Table \ref{t:parameters}  shows the values of the  microturbulence  calculated
for each star. We found a small discrepancy between the  $\xi$$_t$ that we found for CoRoT-4 (1.37~km\,s$^{-1}$)
and that found by \cite{moutou08} (0.94~km\,s$^{-1}$), although both works show an  approximate solar value.
This difference may be caused by the number of iron lines  and the method  used for determining this parameter. 
\cite{moutou08}  used the automatic VWA method \citep{bruntt08}, where they adjusted the microturbulence using an iterative
procedure requiring that there is no correlation between the abundance and strength of a number of weak sensitive Fe I lines with
equivalent widths $<$ 80 m$\AA$.

However, \cite{moutou08} are focused on presenting  the discovery of the CoRoT-4b planet,  and do not present a 
detailed analysis of the abundances values of the elements. We note, nevertheless, that our stellar parameters values,
even if they have larger errors, are not incompatible  with those of \cite{moutou08} in terms  of $T_{\rm eff}$  and log g.

\begin{table}[!]
\caption{Abundances of CoRoT-2 and CoRoT-4}
\label{t:abundances}
\begin{tabular}{l | c c  |c  c c c  }
\hline\hline
Species      &  CoRoT-2       & $\#$ of& CoRoT-4   &   $\#$ of     \\ 
%\hline 
             &  [X/H]         &lines  &  [X/H]     &  lines  \\
\hline
 O\,{\sc i}   & -0.13  $\pm$  0.10       & 3  &   -0.03   $\pm$  0.10     & 3   \\ 
Na\,{\sc i}  & 0.02  $\pm$   0.03     & 3  & 0.05   $\pm$   0.10  & 3 \\   
Mg\,{\sc i}  & 0.04                   & 1  & 0.1                  & 2 \\   
Al\,{\sc i}  & -0.05                  & 2  & -0.02                & 2  \\  
Si\,{\sc i}  & 0.04  $\pm$   0.05     & 4  & 0.01   $\pm$   0.08  & 4  \\  
 S\,{\sc i}   & 0.00     $\pm$  0.09  & 1  &   0.15 $\pm$   0.08  & 1  \\          
Ca\,{\sc i}  & 0.1   $\pm$   0.05     & 7  & 0.1    $\pm$   0.05  & 7  \\  
Sc\,{\sc ii} & -0.04                  & 2  & -0.03   $\pm$   0.05 & 3  \\  
Ti\,{\sc i}  & 0.09  $\pm$   0.09     & 5  & 0.06   $\pm$   0.11  & 3  \\  
Cr\,{\sc i}  & 0.01  $\pm$   0.08     & 4  & 0.1    $\pm$   0.04  & 6  \\  
Mn\,{\sc i}  & -0.06                  & 2  & 0.02   $\pm$   0.03  & 3  \\  
Fe\,{\sc i}  & 0.03  $\pm$   0.05     & 32 & 0.12   $\pm$   0.05  & 27  \\ 
Fe\,{\sc ii} & 0.02  $\pm$   0.07     & 7  & 0.12   $\pm$   0.05  & 7  \\  
Ni\,{\sc i}  & -0.08  $\pm$   0.06    & 6  & 0.17   $\pm$   0.04  & 6  \\  
Zn\,{\sc i}  & -0.1                   & 2  & 0.01      --         & 2  \\  
Ba\,{\sc ii} & 0.19  $\pm$   0.04     & 3  & 0.14   $\pm$   0.06  & 3   \\ 
\hline                                                        
 \end{tabular}                                          
 \end{table}

\begin{table}[[t!]  %Tab 5
\caption{Abundance uncertainties of CoRoT-2.}
\label{t:errorcorot2}
\begin{tabular}{lcccccc}\hline\hline
Species & $\Delta$ T$_{\rm eff}$ & $\Delta$ log g & $\Delta$ $\xi$ &
$\Delta$ W$_{\lambda}$ & $\left( \sum \sigma^2 \right)^{1/2}$ &
$\sigma$$_{\rm obs}$ \\
$_{\rule{0pt}{8pt}}$ & $+70$~K  & $+0.12$      & $+0.10$         & 3 m\AA &   &   \\
\hline
Fe\,{\sc i}    &    0.05    &  -0.01    &   -0.02   &  0.05   & 0.07    &   0.05  \\
Fe\,{\sc ii}   &    -0.02   &  0.05     &   -0.02   &  0.06   & 0.08    &   0.07  \\
Na\,{\sc i}    &    0.04    &  -0.02    &   -0.01   &  0.03   & 0.05    &   0.03  \\
Mg\,{\sc i}    &    0.04    &  -0.03    &   -0.02   &  0.03   & 0.06    &   --    \\
Al\,{\sc i}    &    0.02    &  -0.02    &   -0.01   &  0.03   & 0.04    &   --    \\
Si\,{\sc i}    &    0.01    &  0.01     &    0.0    &  0.04   & 0.04    &   0.05  \\
Ca\,{\sc i}    &    0.05    &  -0.02    &   -0.02   &  0.04   & 0.07    &   0.05  \\
Sc\,{\sc ii}   &    0.0     &  0.05     &   -0.02   &  0.05   & 0.07    &   --    \\
Ti\,{\sc i}    &    0.08    &  0.0      &   -0.01   &  0.05   & 0.09    &   0.09  \\
Cr\,{\sc i}    &    0.06    &  -0.01    &   -0.03   &  0.05   & 0.08    &   0.08  \\
Mn\,{\sc i}    &    0.07    &  0.0      &   -0.01   &  0.05   & 0.09    &   --    \\
Ni\,{\sc i}    &    0.05    &  0.0      &   -0.03   &  0.05   & 0.08    &   0.06  \\
Zn\,{\sc i}    &    0.0     &  0.02     &   -0.03   &  0.05   & 0.06    &   --    \\
Ba\,{\sc ii}   &    0.02    &  0.03     &   -0.05   &  0.05   & 0.08    &   0.04  \\
\hline
Li\,{\sc i}   &     0.06    & 0.00      &   0.01    &  --    &  0.06    &    0.05       \\
 O\,{\sc i}    &     -0.03   & 0.07      &   0.00    & --     &  0.08    &    0.10       \\
 S\,{\sc i}    &     -0.05   &  0.05     &   0.00    &   --   &  0.07    &    0.09    \\

\hline
\end{tabular}
\tablefoot{
 The second column indicates
the variation in the abundance  caused by the variation in
$T_{\rm eff}$. The other columns refer, respectively, to the variations  caused by
to $\log g$, $\xi_{{\rm t}}$, and $W_\lambda$. The sixth column indicates
the combined rms uncertainty of the second to fifth columns. The
last column indicates the observed abundance dispersion for those elements
whose abundances were derived using more than two lines.}
\end{table}                                                      

\begin{table}[t]  %Tab 5
\caption{Abundance uncertainties of CoRoT-4.}
\label{t:errorcorot4}
\begin{tabular}{lcccccc}\hline\hline
Species & $\Delta$ T$_{\rm eff}$ & $\Delta$ log g & $\Delta$ $\xi$ &
$\Delta$ W$_{\lambda}$ & $\left( \sum \sigma^2 \right)^{1/2}$ &
$\sigma$$_{\rm obs}$ \\
$_{\rule{0pt}{8pt}}$ & $+70$~K  & $+0.12$      & $+0.1$         & 3 m\AA &   &   \\
\hline
Fe\,{\sc i}    & 0.05   &  -0.01    & -0.03    &  0.05  & 0.08     &   0.05   \\
Fe\,{\sc ii}   & -0.01  &  0.04     & -0.04    &  0.05  & 0.08     &   0.05   \\
Na\,{\sc i}    & 0.04   &  -0.01    & 0.0      &  0.05  & 0.06     &   0.10   \\
Mg\,{\sc i}    & 0.04   &  -0.01    & -0.01    &  0.04  & 0.06     &   --   \\
Al\,{\sc i}    & 0.03   &  -0.01    & 0.0      &  0.05  & 0.06     &   --   \\
Si\,{\sc i}    & 0.03   &  0.0      & -0.01    &  0.05  & 0.06     &   0.08   \\
Ca\,{\sc i}    & 0.04   &  -0.02    & -0.03    &  0.04  & 0.07     &   0.05   \\
Sc\,{\sc ii}   & 0.0    &  0.04     & -0.03    &  0.05  & 0.07     &   0.05   \\
Ti\,{\sc i}    & 0.06   &  -0.01    & -0.01    &  0.08  & 0.10     &   0.11   \\
Cr\,{\sc i}    & 0.06   &  -0.01    & -0.04    &  0.06  & 0.09     &   0.04   \\
Mn\,{\sc i}    & 0.06   &  0.0      & -0.01    &  0.06  & 0.09     &   0.03   \\
Ni\,{\sc i}    & 0.06   &  -0.01    & -0.04    &  0.05  & 0.09     &   0.04   \\
Zn\,{\sc i}    & 0.03   &  0.0      & -0.05    &  0.06  & 0.08     &   --   \\
Ba\,{\sc ii}   & 0.02   &  0.02     & -0.08    &  0.05  & 0.10     &   0.06   \\
\hline                                                    
Li\,{\sc i}   & 0.05    &  0.00      &  0.00    &  --    & 0.05   &  --         \\
O\,{\sc i}    & 0.12    &  0.07     &  0.08    &  --    & 0.16   &    0.10       \\
S\,{\sc i}    & -0.03   &  0.05     &  0.00    &  --    & 0.06   &      0.08       \\
\hline                                                        
\end{tabular}
\tablefoot{ The second column indicates
the variation of the abundance  caused by the variation in
$T_{\rm eff}$. The other columns refer, respectively, to the variations caused by
$\log g$, $\xi_{{\rm t}}$, and $W_\lambda$. The sixth column indicates
the combined rms uncertainty of the second to fifth columns. The
last column indicates the observed abundance dispersion for those elements
whose abundances were derived using more than two lines.}
\end{table}

\subsection{Lithium}

Of the two observed stars, only CoRoT-2 exhibits a relatively strong lithium resonance line
with a measured equivalent width of 124~m\AA. The secondary  Li\,{\sc i} line at 6104~\AA\
appears to be present in the spectrum as a very small feature in the wing of a Fe\,{\sc i}
line, confirming indirectly the high  lithium abundance of this star. The lithium abundances
were derived from synthetic spectra matching the observed  Li\,{\sc i} 6708~\AA\ resonance doublet.

The wavelengths and oscillator strengths for the individual hyperfine and isotope
components of the lithium lines were taken from \cite{smith98} and \cite{hobbs99}.
The CN lines in the vicinity of the Li\,{\sc i} doublet were included in the linelist.
The solar $^6$Li/$^7$Li isotopic ratio ($^6$Li/$^7$Li$=0.081$) \citep{anders89} was
adopted in the synthetic spectrum calculations.

\begin{figure}
\resizebox{\hsize}{!}{\includegraphics{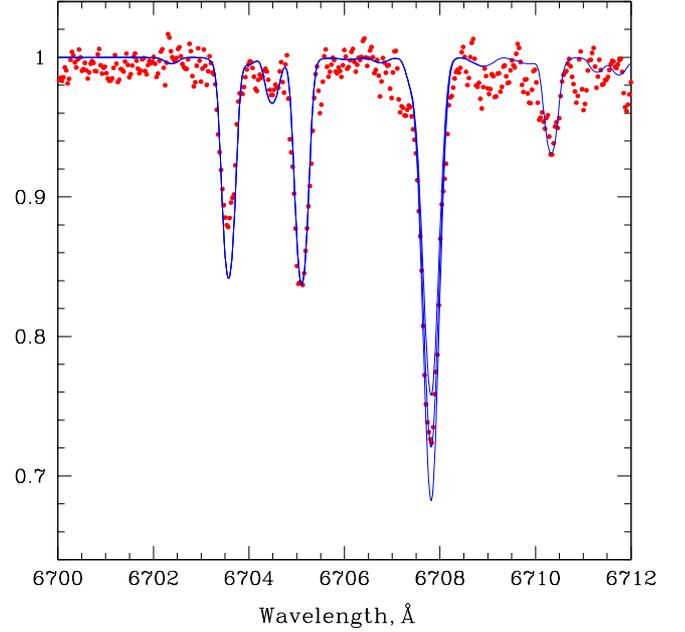}}
\caption{CoRoT-2: Observed (dotted red line) and synthetic (solid blue lines)
spectra in the region of the Li\,{\sc i}  6708~\AA\ line. Synthetic
spectra were used to calculated the lithium abundances (from top to bottom):
$\log\varepsilon{\rm (Li)}=$ 2.6, 2.7, and 2.8,.
 The best-fit function to the observations is achieved  for lithium
abundance $\log\varepsilon{\rm (Li)}= 2.7$  $\pm$ 0.05. We also show three iron absorption lines.
}
 \label{f:li-c2}
\end{figure}

For CoRoT-2, we obtained $\log\varepsilon({\rm Li}) = 2.70  $  $\pm$ 0.05.
The observed and synthetic spectra of CoRoT-2 in the region of Li\,{\sc i} line
are shown in Fig.\ref{f:li-c2}. For the secondary Li~{\sc i} line, we only obtained the upper limit to the Li
abundance of $\log\varepsilon{\rm (Li)}$$<$ $2.8$, which agrees with that obtained for the
resonance line. The high Li abundance determined  for  CoRoT-2 of effective temperature
$T_{\rm eff}$ = 5696~K and both its chromospheric activity and
relatively high rotation velocity v$\sin i=8.5$~km\,s$^{-1}$ imply a young
age of this star.

\begin{figure}
\resizebox{\hsize}{!}{\includegraphics{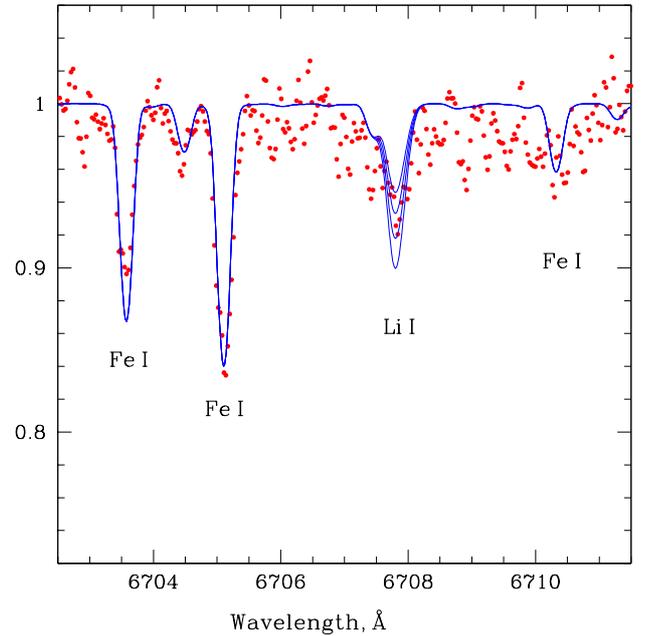}}
\caption{CoRoT-4: Observed (dotted red line) and synthetic (solid blue lines)
spectra in the region of the Li\,{\sc i}  6708~\AA\ line. Synthetic
spectra were used to  calculated  the lithium abundances (from top to bottom):
$\log\varepsilon{\rm (Li)}=$ 2.1, 2.2, 2.3 and 2.4. It is  difficult to determine
a value for the abundance because of the noisy spectrum, and we  only estimated an upper
limit of $\log\varepsilon{\rm (Li)}= 2.2$.  We also show three iron absorption lines.
 }
 \label{f:li-c4}
\end{figure}

The star with the higher effective temperature
$T_{\rm eff}$= 6115~K, CoRoT-4, exhibits a relatively weak Li\,{\sc i} resonance line.
Analyzing the relatively noisy spectrum of this star in Fig.\ref{f:li-c4}, we were unable to obtain a reliable 
value of the Li abundance. For CoRoT-4, we estimate only an upper limit to the lithium abundance on  the order of
$\log\varepsilon{\rm (Li)}$ $<$ 2.2. This apparently high  Li abundance of CoRoT-4 is, however,
consistent with the Li abundance of the main-sequence field and cluster stars
with effective temperatures of $5900<T_{\rm eff}< 6300$~K
(the so-called Li-plateau region according to \cite{boesgaard04}.)

%nova

%nova

\subsection{Oxygen}

To determine of oxygen abundance, we used the O\,{\sc i}
triplet lines at 7771, 7774, and 7775 ~\AA\ .
The atomic data were taken from the VALD database \citep{piskunov95,kupka99}.
The value of the oscillation strength of the main component was modified to fit the solar
abundance of $\log\varepsilon({\rm O})=8.93$ (Anders \& Grevesse 1989).

The oxygen triplet lines are known  to be affected by NLTE effects, which lead to an overestimation of the oxygen
abundances. Even if the forbidden [O\,{\sc i}] line at 6300.304~\AA\ is assumed to be
free from NLTE effects,  this line is blended with a telluric [O\,{\sc i}]
emission line in our spectra and cannot be used for the oxygen abundance determinations. To account for
NLTE effects, we used the theoretical work of \cite{takeda03}, who calculated the values of
the NLTE corrections for each line of the O\,{\sc i} infrared triplet for a grid of
different atmospheric parameters and equivalent widths of the oxygen lines.

For the cooler star of our sample, CoRoT-2, the NLTE correction is  about 0.1~dex,
whereas for the hotter star, CoRoT-4, NLTE effects are greater,  about 0.2~dex.
By applying the NTLE corrections, we derived the  oxygen  abundances
$\log\varepsilon({\rm O})=8.80$ and 8.90 for CoRoT-2 and CoRoT-4, respectively.

\begin{figure}[t]                                                                                              
\resizebox{\hsize}{!}{\includegraphics{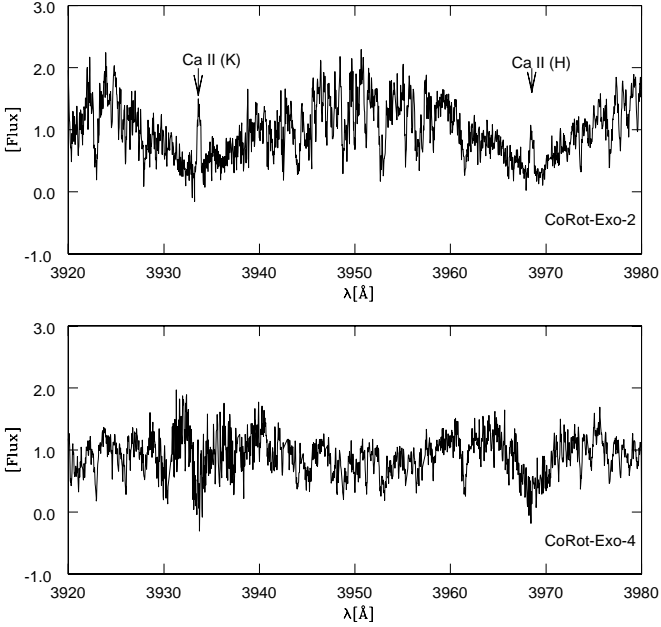}}                                                     
\caption{Calcium II K and H lines for stars CoRoT-2 and CoRoT-4. The central emission
features in the Ca lines correspond to the younger and chromospheric active CoRoT-2 star. }
 \label{f:ca}                                                                                                
\end{figure}

For CoRoT-2, not only  a strong resonance Li~{\sc i} line but also 
emission in the Ca\,{\sc ii} lines is a strong indication of its
youth and surface  activity (see Fig.  \ref{f:ca}). 
In the case of CoRoT-4,  emission is absent (Fig.  \ref{f:ca}). This age-related
activity is proposed  to affect the oxygen abundances derived from the oxygen
triplet lines \citep{shen07} and may also affect the precision of the oxygen
abundance determination for CoRoT-2 star.

\subsection{Sulfur}

Sulfur is  a very important volatile element, being one of the commonly and
observed elements, the only reference element in an extended interval of $T_{\rm C}$
values between 704~K and 371~K \citep{lodders03}. To obtain the S abundances in both stars, we selected the line of S\,{\sc i}
at 6757~\AA. This line consists of three fine structure components and has to be 
studied using the synthetic spectrum method. We used the $\log gf$ data of \cite{wiese69}, which were provided by VALD. 

The value of the oscillation strength of the main component  (--0.310) was modified to --0.340 to fit the solar
abundance of $\log\varepsilon({\rm S})=7.21$ (Anders \& Grevesse 1989). The sulfur abundances that we derived are
$\log\varepsilon({\rm S})=7.21 \pm 0.09$ and $\log\varepsilon({\rm S})=7.36 \pm 0.08$ for CoRoT-2 and CoRoT-4,
respectively.

\section{Numerical simulations:analysis and methodology}

\subsection{ The planetesimal disk}

A protoplanetary disk contains a mixture of gas and dust grains. These dust grains collide and build
up larger and larger bodies, and small solid bodies known as planetesimals are presumably  formed.  Protoplanets
grow within protoplanetary disks via pairwise accretion of planetesimals.  When the mass of a protoplanet
is of the order of 10-20 terrestrial masses, it can gravitationally capture gas-rich disk, thus producing Jovian-type
planets. Protoplanetary disk is expected to lose its gas component in the first 10 Myr, leaving a disk formed
essentially of dust and planetesimals.

For our purposes, we assumed that large planets observed close to their parent stars actually
formed at larger distances but migrated inward due to energy loss and  excess energy used to disperse
the planetesimals \citep{murray98}.  A detailed description of the planet migration used in this
work is presented in \cite{winter07}. It consist of a forced inward migration of a planet supposedly
interacting with a planetesimal disk. Several possibilities were considered in terms of the planet
eccentricities, migration rates, and mass. A more general result from \cite{winter07} is that 2:1 mean motion
resonance is the main mechanism for driving planetesimals to the surface of the star. This mechanism operates
for slow migration rates ($\tau=10^{5} - 10^{6}$ yr) and low planet eccentricities.

To construct a rocky disk model valid for ages equal to or older than 10 Myr, several considerations must
be taken into account. This disk containing a planetesimal distribution can be modeled using, for instance,
knowledge obtained about a solar primeval nebula. Based on \cite{lodders03}, in Table  \ref{t:lodder} we present  the distribution
of the most important and more abundant rock-forming elements. We indicate the type of elements (from highly refractory up to highly
volatile), their proper elements, their interval of condensation temperature,  and finally,
their semi-major axis  distribution (in AU) with respect to a central solar type star whose
radial thermal distribution is valid for an age older than 10 Myr.

\begin{table}[b]
\begin{center}
%\centering
\caption{Most important and more abundant rock-forming elements (from \cite{lodders03})}             % title of Table
\label{t:lodder}
\begin{tabular}{|l| c| c| c|}        % centered columns (4 columns)
\hline
Type of element     &     Element              &   $T_{\rm C}$ ($^o$K )  & a (AU )      \\
\hline
Refractory          & Ca,  Al,  Ti             & 1650 - 1360   &  0.03 - 0.04   \\
Common              & Fe, Si, Mg               & 1360 - 1290   &  0.04 - 0.046  \\
Moderately Volatile &        P                 & 1290 - 704    &  0.046 -  0.15 \\
Volatile            & S                        & 704 - 371     &  0.15 - 0.48   \\
Highly Volatile     & H, C, N, O +   &    $<$ 371      &  0.48 - $<$ 5    \\
                    & He, Ne, Ar               &               &                    \\
\hline
\end{tabular}
\end{center}
\end{table}

For simplicity, we assume that in a certain epoch of the disk evolution, most probably
in the range  20 -- 30 Myr (see Sect. 5), the disk is formed by a sea of planetesimals
defining three representative zones: a  refractory  rich planetesimals  zone R between
$\sim$ 0.03 AU and $\sim$ 0.1 AU (between 1780~K and 1360~K), an  intermediate zone I for
particles between $\sim$ 0.1 AU and 1.56 AU (1360~K to 200~K), and a zone V of  volatiles
between 1.56 AU and $\sim$ 4.5 AU ( $<$ 200~K).

In reality, the situation must be far more complex. Because  nebular turbulence and dust
migration probably occur during the earlier nebular stages, a radial mixture is expected.
If zone {\it R} contains  the pure refractory elements, zone {\it I} must contain a true 
mixture of elements. An example of this mixture in zone {\it I}, is the presence of
Triolite (FeS) at 0.15 AU with a $T_{\rm C}$ 700 K \citep{lodders03}. At this distance,
an important condensation of pure Fe grains and a mild condensation of pure C may also
occur \citep{wehrstedt02}. The zone {\it I} also contains the internal end of the migration
condensation front of H$_2$O particles at $\sim$ 0.8 AU coming from the external zones 
of the disk \citep{davis05}. In summary, we can consider that zone {\it I} contains
planetesimals composed of a mixture of refractory type grains and volatile ices.
The zone V  is characterized by the abundant highly volatile CNO elements for which
the highest  $T_{\rm C}$ is only 180~K for oxygen.

\subsection{The planet migration}

We numerically integrated the  CoRoT-2, CoRoT-3, and CoRoT-4 systems within the circular
restricted three-body problem of star-planet-planetesimals. The orbital and physical parameters
for these stars are listed in Table  \ref{t:6}. Each parameter is listed along with its formal
uncertainty. We assumed that the planet may have been formed  at a distance between  2 and 5 AU,
where they are supposed to form under conditions where condensible materials exist.
Thus,  the planet is set initially with semi-major axes $a_{P_i}=$2, 3, 4, and 5 AU and eccentricity $e_{P}=0.0$.
The planet is forced to migrate inward  to its current position, $a_{P_f}$, with constant speed in a time-scale $\tau$.
We assumed that the planetesimal disk is composed of 1000 test massless particles moving around the star on
initially randomly distributed circular orbits. The disk is interior to the orbital radius of the planet.
The newly formed planet is  surrounded by an annular gap in the planetesimal disk
 \citep{wisdom80,lissauer93}. We assumed here  an annular gap of  0.5 AU relative to the outer disc edge.
Therefore, the outer radius is, in all cases, 0.5 AU smaller than the initial position of the planet.
Thus the disk outer radius ($a_{D_f}$) is located at 1.5, 2.5, 3.5, and 4.5 AU, according to the planet
initial semi-major axis. The inner radius ($a_{D_i}$) of the disk is located at 0.03 AU.

We are interested in studying the relationship between the metallicity of these CoRoT systems and
the migration resulting from interaction with planetesimals. Since a part of the mass of planetesimals
will be accreted by the star during migration,  we need to estimate for these CoRoT systems  the mass 
of the disk contained inside the orbit of each planet. Considering the initial semi-major axes of the
planets, we calculated the orders of the total mass $M_D$ of the planetesimal disk required for the
planet to migrate  to its current position \citep{adams03}.
These data are presented in Table  \ref{t:7}. We assumed that about 1$\%$ of the disk is in the form of solid
bodies called planetesimals \citep{davis05}. We note that the order of the total mass of the
planetesimal disk increases with  planetary  mass and larger $a_{P_i}$. Using these values, we now
estimate the percentage of the element types that fall on the star during the planet migration.
The numerical code SWIFT \citep{levison94} was used. The integrator is based on the MVS method developed by
\cite{widsom91}.

\begin{table*}
\centering
\caption{Orbital and physical parameters of the stellar systems}
\label{t:6}
\begin{tabular}{|l| c| c| c| c |c|}        % centered columns (4 columns)
\hline

Orbital parameter     & M$_{star}$ (M$_{\odot}$) & M$_{planet}$ (M$_{Jupiter})$ & Semi-major axis (AU)  & Eccentricity & References   \\
\hline
Corot-2b &   0.97 $\pm$  0.06          & 3.31 $\pm$  0.016  &   0.0281 $\pm$ 0.0009   & 0.0 (fixed) & \cite{bouchy08}\\
Corot-3b &   1.37 $\pm$  0.09          & 21.66 $\pm$  1.0   &   0.057  $\pm$ 0.003    & 0.0  (fixed) & \cite{deleuil08} \\
Corot-4b &   1.16 $_{-0.02}$$^{+0.03}$ & 0.72  $\pm$  0.08  &   0.09   $\pm$ 0.001    &  0.0$_{-0.0}$$^{+0.1}$  & \cite{moutou08} \\
                   
\hline
\end{tabular}

\end{table*}

\begin{table*}
\centering
\caption{Total mass of planetesimals for the planet migration up to its current position  in the  stellar system}
\label{t:7}
\begin{tabular}{|l| c| c| c|}        % centered columns (4 columns)
\hline
 Disk mass ($M_D$)    &   Corot-2b &   Corot-3b  & Corot-4b      \\
\hline
 $a_{P_i}$ (AU)  & 2 - 5 & 2 - 5  & 2 - 5 \\
 \hline
$M_{Disk}$ ($M_{\bigodot}$) &  0.013 - 0.016  &   0.075 - 0.093   & 0.002 - 0.003   \\

\hline
\end{tabular}

\end{table*}

In general, the numerical simulations were  performed  for a timescale of $10^5$ years.  
The integration for each particle was interrupted whenever one of the following
situations occurred:

(i) collision between the planetesimal and the planet, i.e., when the planetesimal becomes closer 
than 1.2 planetary radii from the planet;

(ii) collision between the planetesimal and the star, when the planetesimal becomes closer 
than 2.5 stellar radii from the star;

(iii) ejection from the system, when the planetesimal reaches more than 50 AU from the star.

\subsection{ Results}

\begin{figure*}[!]
\begin{center}
\includegraphics[width=0.3\textwidth]{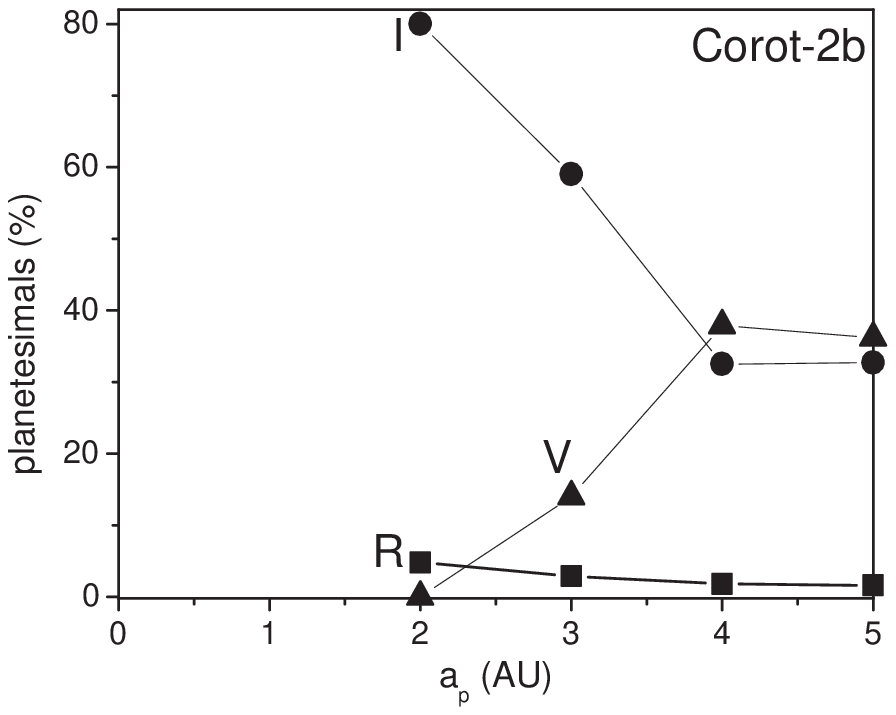}
\includegraphics[width=0.3\textwidth]{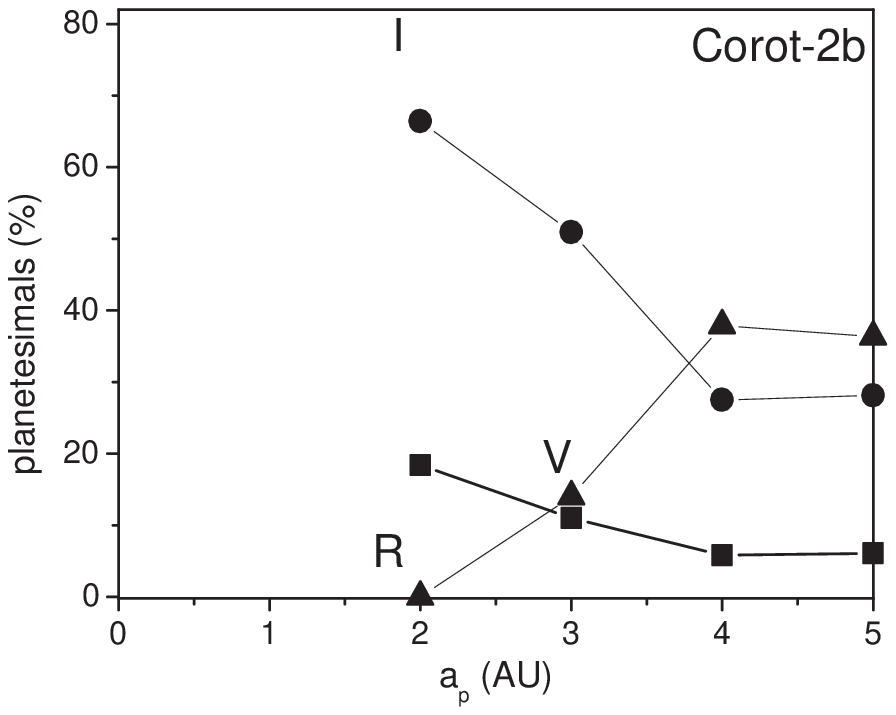}
\includegraphics[width=0.3\textwidth]{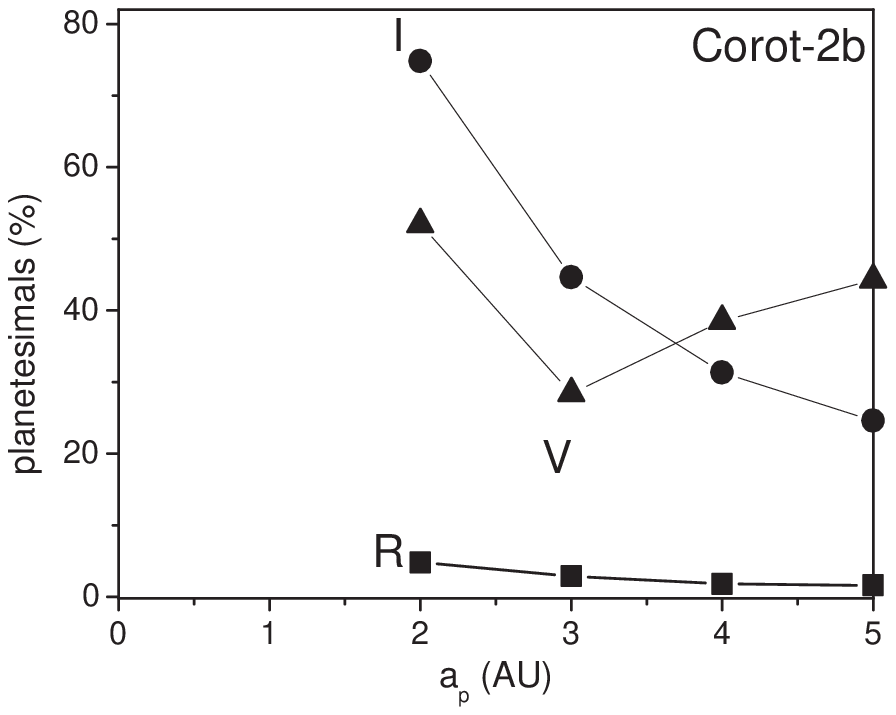}
\includegraphics[width=0.3\textwidth]{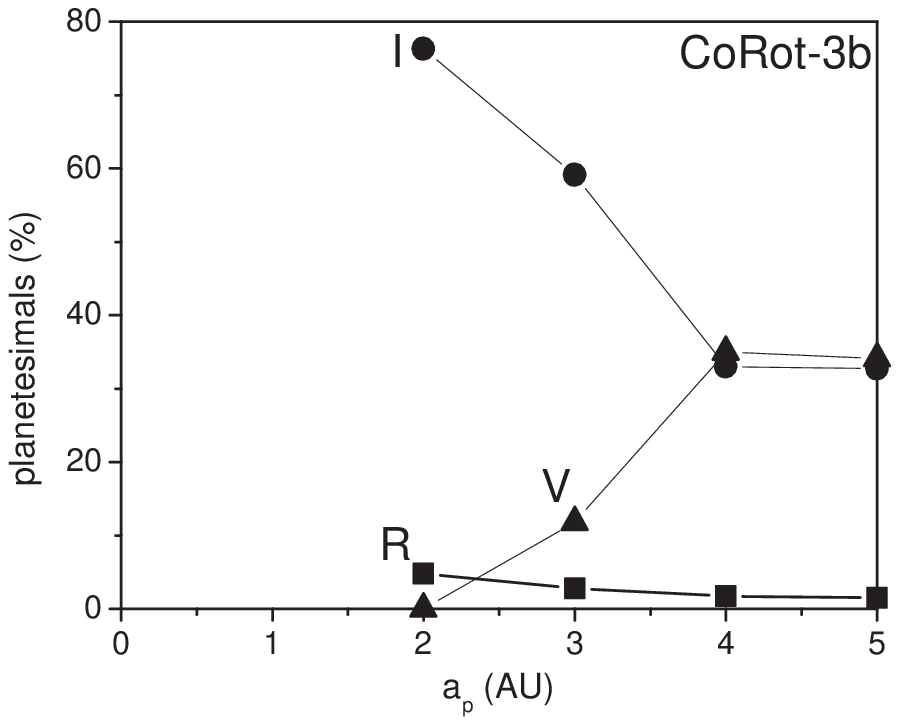}
\includegraphics[width=0.3\textwidth]{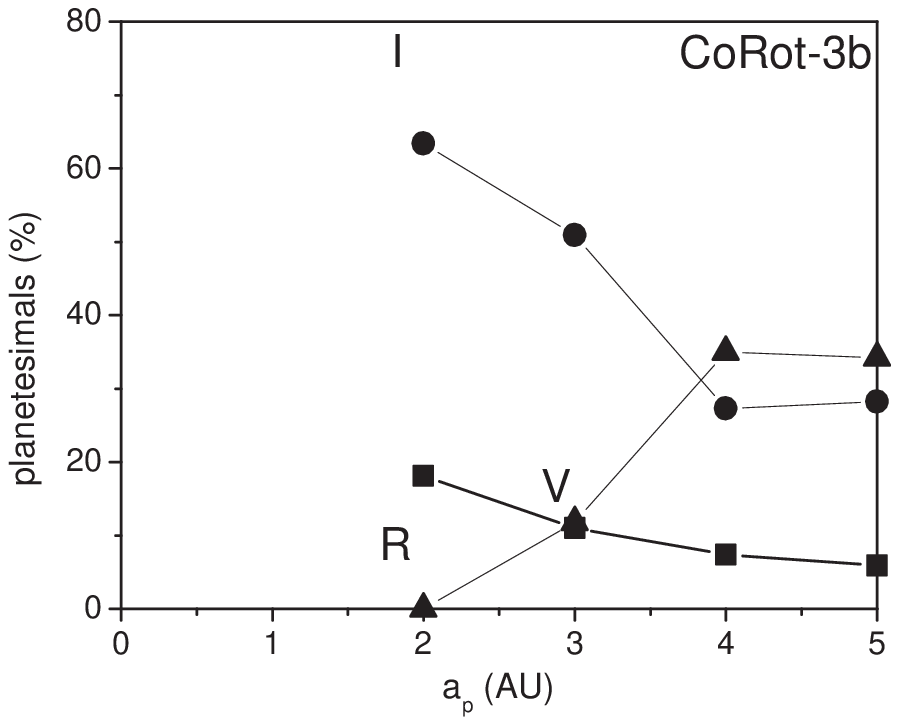}
\includegraphics[width=0.3\textwidth]{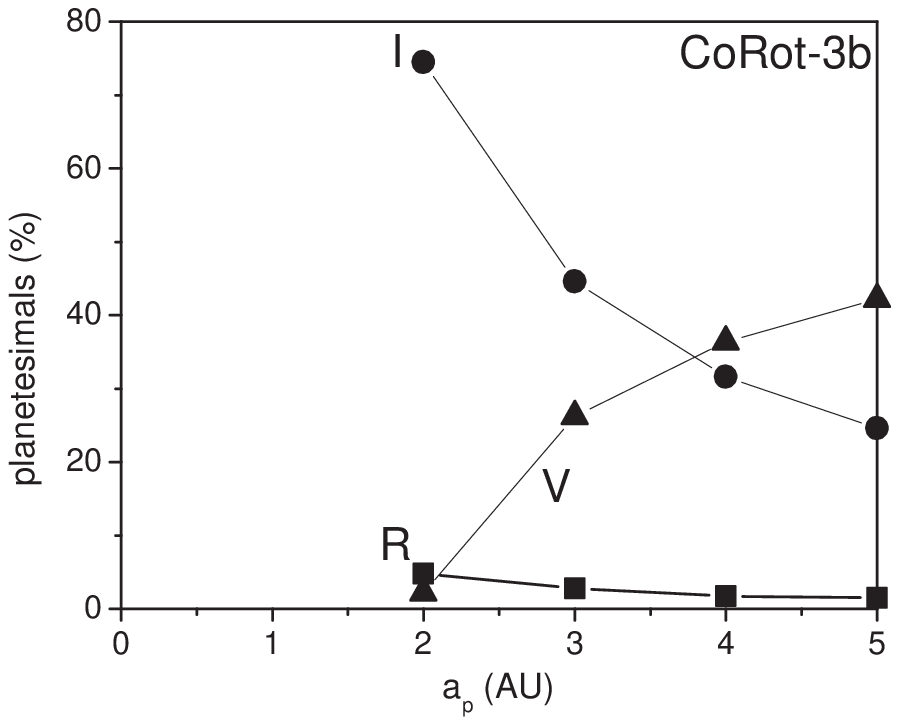}
\includegraphics[width=0.3\textwidth]{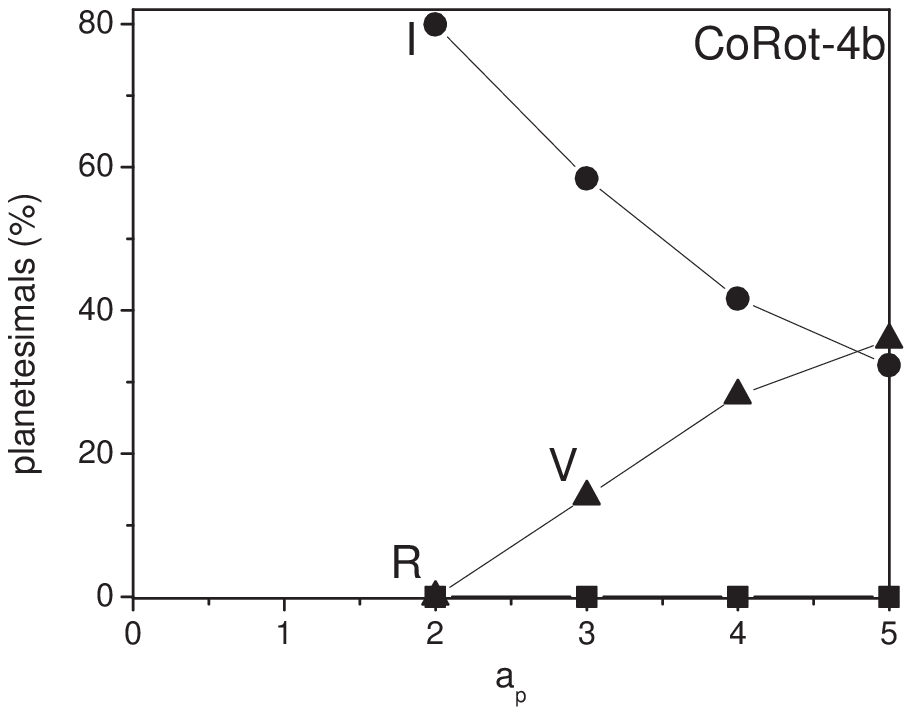}
\includegraphics[width=0.3\textwidth]{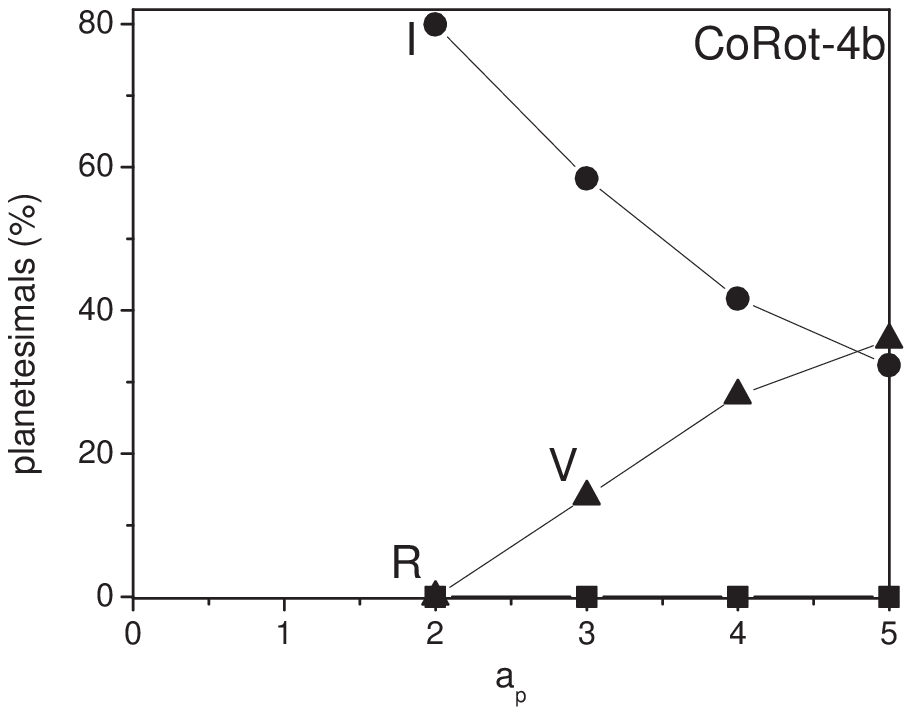}
\includegraphics[width=0.3\textwidth]{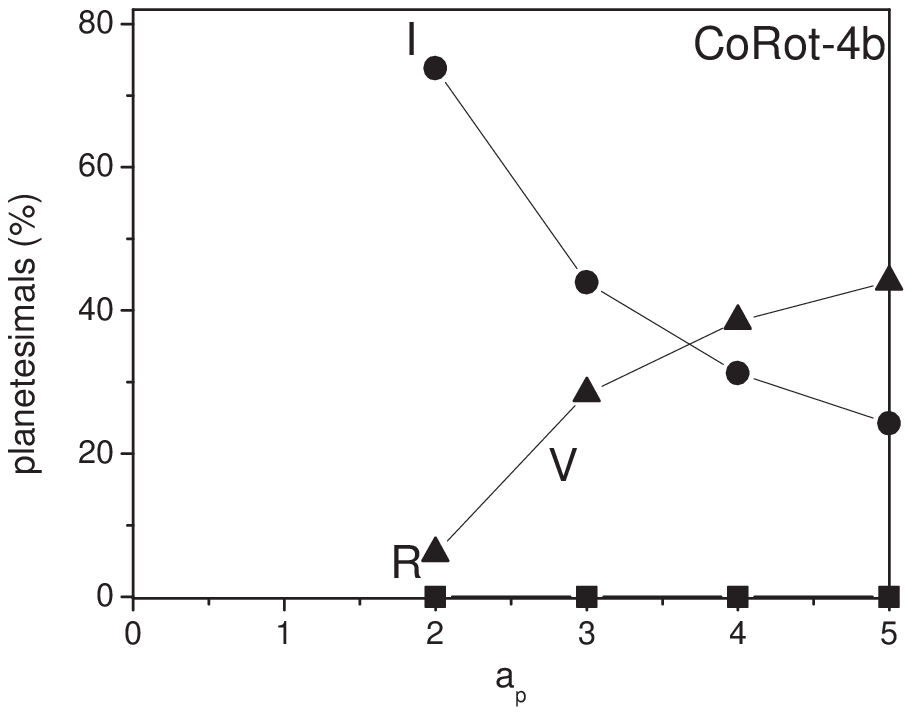}
 \caption{\label{f:4} These figures show the percentage of planetesimals that collided with the star of the
 CoRoT systems. Each curve gives the percentage  of planetesimals of the zones refractory (square), intermediate
 (circle) and volatile (triangle). The values are given as a function of the initial semi-major axis of the planet,
 $a_{P_i}$. At the top of each figure, we indicate the corresponding system. In the first column, the results are for the
 case of zones R between 0.03 AU and 0.1 AU,  I  between 0.1 AU and 1.56 AU, and  V between 1.56 AU and 4.5 AU. The
 second column is for the case of zones R between 0.03 AU and 0.3 AU,  I  between 0.3 AU and 1.56 AU, and  V between
1.56 AU and 4.5 AU. The third column is for the case of zones R between 0.03 AU and 0.1 AU,  I  between 0.1 AU and 1.2 
AU, and  V between 1.2 AU and 4.5 AU.}
\end{center}                                                                                                                                     
\end{figure*}

The results of the numerical simulations are presented in Fig.  \ref{f:4}. In that figure, we
show the  dependence of the percentage of planetesimals that fall on the star in terms of the value of $a_{P_i}$.
The main significance of the results are the statistics  in terms of percentage of collisions with the star of
planetesimals characterized by R, I, and V as a function $a_{P_i}$. The values of the percentages were obtained by
measuring the number of planetesimals of the  R, I, and V zones that collided with the star.
Thus the percentage of collisions of planetesimals with the star gives an estimate of the mass of each type
of material that the star will accrete due to the migrating planet. We note that while this percentage provide
reasonable estimates  of the accreted mass, the assumption of there being a constant migration is not strictly
correct since these collisions will tend to decrease as the planet migrates inwards.

We can also note from Fig. \ref{f:4} that the percentage of planetesimals in the volatile zone that
fall on the star is increases with increasing  value of $a_{P_i}$. This is due to the disk model considered.
 As mentioned in the previous section, the circumstellar disk is assumed to contain highly volatile particles between
1.56 and 4.5 AU. They are distributed in that region for values of the initial position of the planet beyond  2 AU.
Thus,  there are fewer particles in the volatile zone when the planet has an initial position 2 AU.
In this case, all the particles are practically initially distributed in refractory and intermediate zones. Since the size of the
refractory zone is much smaller (0.03 -- 0.1 AU) than the size of the intermediate zone (0.1 -- 1.56 AU),  there is
a much larger quantity of particles in the intermediate zone. 
A more significant number of particles is then expected to fall on the star from the intermediate zone.
Therefore, our results show that for these CoRoT systems,  particles from the intermediate zone
dominate the collision process onto the star.
In total  about  79$\%$  of planetesimals from the intermediate zone were found to be falling onto the star. In the CoRoT-2 and
CoRoT-3 cases, all material ($\sim$ 5$\%$)  of planetesimals from the refractory zone fell onto the star.

For all the migration cases of the planet of the CoRoT-4 system, we note that  planetesimals
from the refractory zone did not fall onto the star. This occurred because the planet migration  halts before the 2:1
resonance reaches  the refractory zone.

When planets migrated from the initial position at 3 AU,  about 3$\%$ of the planetesimals from the refractory zone fell
onto CoRoT-2 and CoRoT-3 stars. Planetesimals from the volatile and intermediate  zones fell on the stars.
The results show that  about 13$\%$ belongs to the volatile zone, while 59$\%$ belongs to the intermediate zone.
For the initial positions of the planet at 4 and 5 AU, our results do not show significant differences between
the values of  percentages of planetesimals from the refractory, intermediate, and volatile zones that fall on CoRoT-2
and CoRoT-3. In these two cases, the numerical estimates are very close.  There is  doubtless no dependence of these
percentages values with the initial semi-major axis of the planet.
In general, giant planets  are, therefore,  be expected to form at distances between 4 and 5 AU. We  focused on the
situation of an orbital radius  at 5 AU, where giant planets form and gain most of their mass.

\begin{figure*}[!]
\begin{center}
\includegraphics[width=13cm]{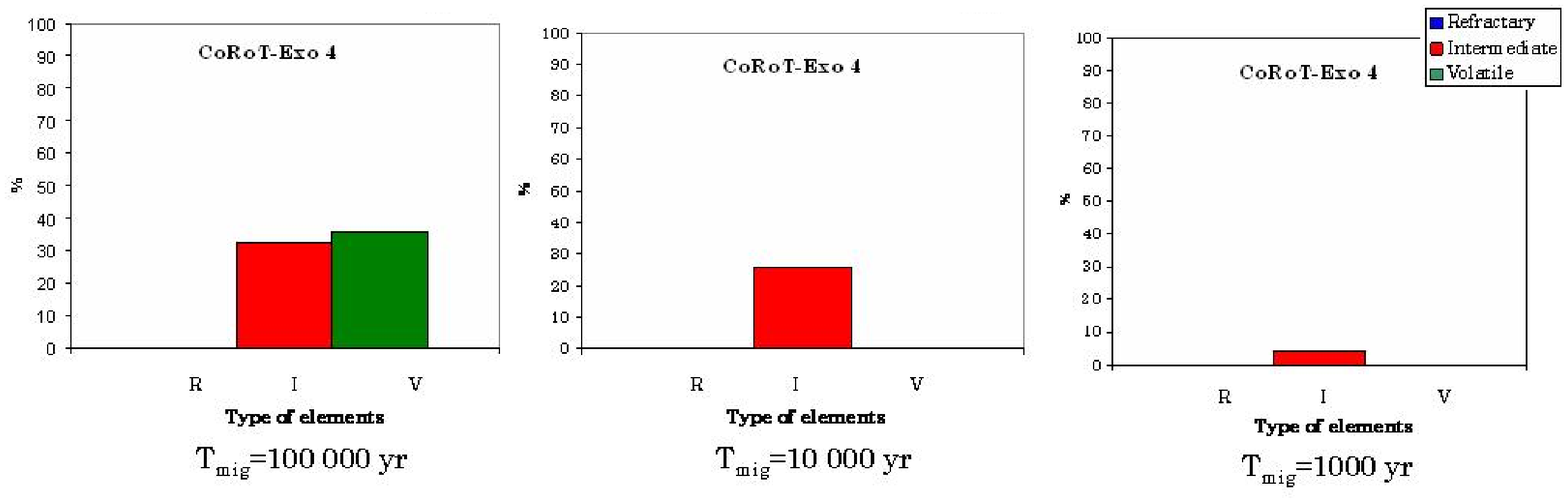}
 \caption{\label{f:histo}\footnotesize{Histogram of the percentage of planetesimals from the refractory (R), 
intermediate (I), and volatile (V) zones that fall on Corot-4 for the case of $a_{P_i}=5$ AU.}}
\end{center}
\end{figure*}

Motivated by the much smaller total mass of its planetesimal disk than the other systems (Table  \ref{t:7}),  the
numerical simulations for the CoRoT-4 case were also  performed for timescales of $10^3$ and $10^4$ years. For very rapid
migrations, the planetesimals experience close approaches to the planet and a significant number of them are ejected from
the system \citep{winter07}. The question that we wish to answer is: how much mass from the disk will the star accrete
during these times?    The results  are shown in Fig. \ref{f:histo}. In this case,  we can note that 26$\%$ of
planetesimals from the intermediate zone  collided with star for $\tau=10^4$ years, while  for $\tau=10^3$ years the 
value is about 3.9$\%$.

\subsection{Parameters dependence}

Until now, the results presented are concerning a given set of  parameters model.  In order to gain more insight into 
the  numerical results  from our disk  simple model, it is useful to characterize and,  when possible, quantify its dependence 
with respect to some of the parameters used.  To investigate the sensitivity of our results, in this section we present
a study about the influence of some of the adopted parameters on the amount of planetesimals that fall on the star.

\subsubsection{R, V and, I location zones}

A starting point for our discussion is provided by the distribution model of planetesimals  in the disk,
which is given by the location of the R, I, and V zones. This has important consequences. It  implies that the percentage 
of the type of elements that fall on the star seems to be uniquely controlled by the location zones considered.
As mentioned in  Section 4.1, in  reality a radial mixture of bodies of zones R and V must be contained in the whole disk.
The intermediate zone is characterized by this mixture. Thus, we studied the sensitivity of our results  to the
location of the zones considered. It is therefore important to know how different locations of the R, I, and V zones
affect the  percentages of collisions of planetesimals with the star. To test the dependence of the results on the location zones,
we relaxed the borders of the locations by considering 2 cases. In the first case, we moved the border between refractory and
intermediate from 0.1 AU to 0.3 AU. In the second case, we moved the border between
 intermediate and volatile from 1.56 AU to 1.2 AU.

Our results suggest that there are significantly change in the   percentage values for the CoRot-2 and CoRot-3 systems
 (see second and third columns of Fig. \ref{f:4}). In these cases, it can be seen that a combination of  higher and lower
  values of the initial distribution borders  R and I (or I and V)  results in  an abrupt transition between refractory  and
 intermediate elements (or intermediate and volatile elements) occurring at the positions that we defined as the borders. Thus,
 a reasonable fraction of particles near the borders change from ``intermediate'' to  ``refractory'' elements (or from
  ``intermediate''  to   ``volatile''  elements). In the first case, the percentage of refractory material that fall  onto
 the star is increased by a fraction and the intermediate material that fall on the star is decreased by that same fraction.
 The same idea is valid for the second case,  a fraction of intermediate material near the border is classified as volatile
 material. Then, the percentages of intermediate and volatile material are increased and decreased by the same amount,
 respectively.

For CoRot-4 system,  there is only a change in the percentage values when the border between intermediate and
 volatile material is moved from 1.56 AU to 1.2 AU. There is no change in the percentage of refractory and volatile materials
 that fall onto the star when the border between refractory and intermediate zones is moved. This is because the
 mean-motion resonances  play the main role in the orbital evolution of the planetesimals and their locations are
 unchanged by the shift of the borders. In this scenario, there is no change in the distribution of the initial orbital
 semi-major axis of the particles, but only a determination of the new location of the R,  V, and I zones.

\subsubsection{Planet on eccentric orbit}

Another point for our discussion is provided by the  planet's eccentricity  $e_P$.
We recall that the value of $e_P$ for any extrasolar planet can vary significantly due
to uncertainties in the current observational methods of  detection. As we can see in Table
\ref{t:7}, the  value of $e_P$ for CoRot-3b and CoRot-2b were assumed to be zero  \citep{alonso08,deleuil08}.
Thus, it is important to know whether different values of $e_P$ yield different evolutions.
\cite{winter07} presented a numerical study about the influence of  $e_P$ on capture of
planetesimals in mean motion resonances during the planet migration. They showed 
that for $M_{planet}=M_{Jupiter}$,  when $e_{P}=0$,  the planetesimals are caught on  3:2, 
5:3, or 2:1 resonances. The 2:1 resonance is responsible for the majority of impacts on the star.
For $e_{P}=0.1$, the 5:2 and 3:1 are also important, but the 2:1 resonance is  strong enough
to dominate the orbital evolution of the planetesimal. In the case $e_{P}=0.2$, the 4:1 resonance
is important, while the 3:2 and 5:3 resonances contribute very little to the evolution of the 
planetesimals. Other resonances also become  important for the cases $e_{P}=$ 0.3, 0.4, and 0.5. 
Therefore, the orbital evolution of the disk of planetesimals is affected by several resonances
 but  not dominated by any particular one.
In these cases, the planetesimals with semi-major axis larger than 0.65 AU are spread without being captured in resonance.
Until  now, we have assumed that the CoRot systems have $e_{P}=0$. A comparison of Fig. \ref{f:4} with the 
results of Fig. 6 from  \cite{winter07} for $e_{P}=0$ and $\tau = 10^5$ yr  show that, in general, the 
percentage of planetesimals that fall on the star  is between about of 70 and 80 $\%$.  \cite{winter07} 
demonstrated that this result is a direct consequence of the value of the migration speed.

\begin{figure*}[!]
\begin{center}
\includegraphics[width=0.3\textwidth]{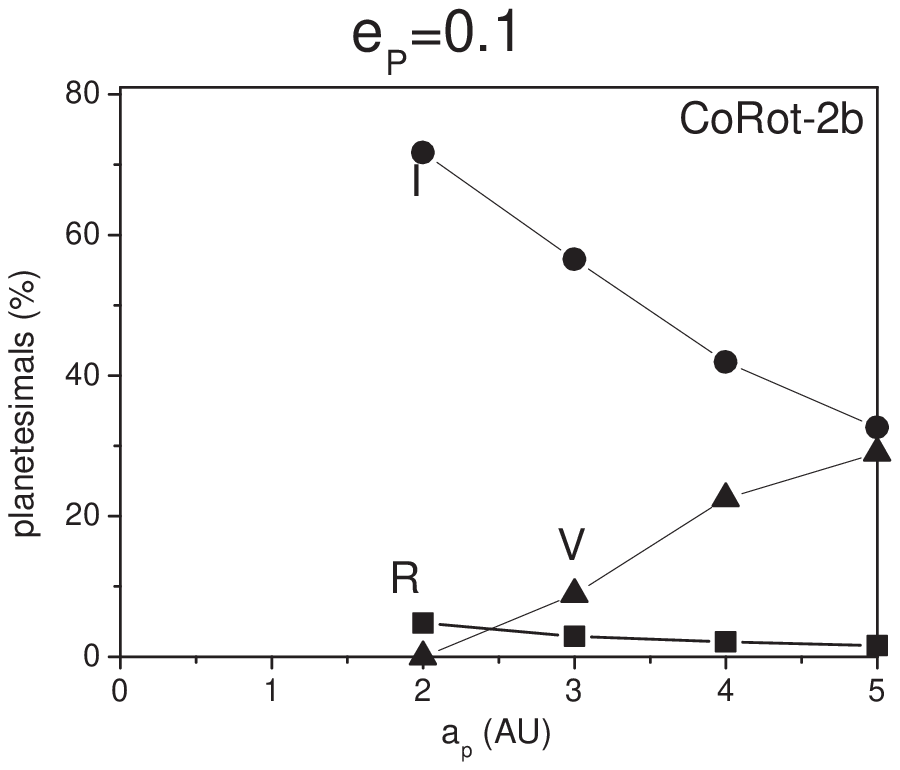}
\includegraphics[width=0.3\textwidth]{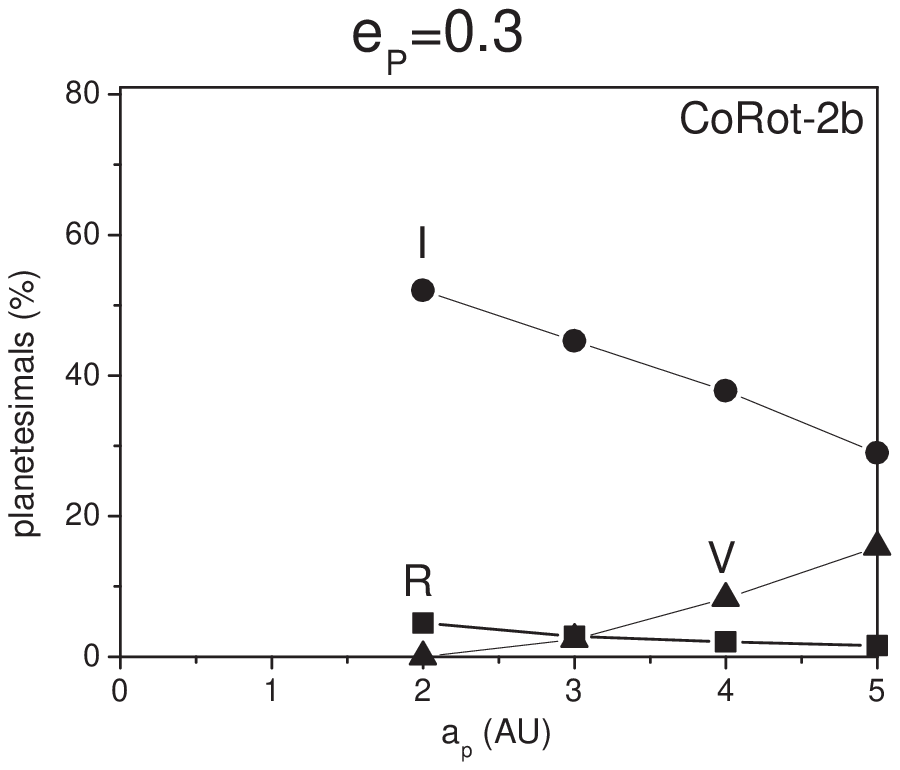}
\includegraphics[width=0.3\textwidth]{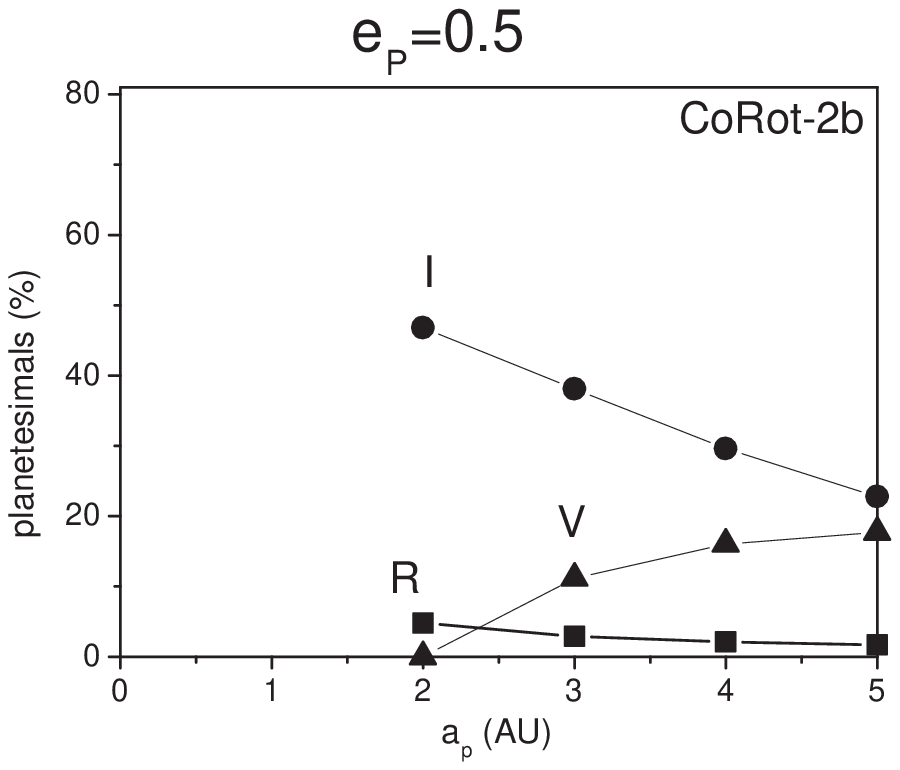}
\includegraphics[width=0.3\textwidth]{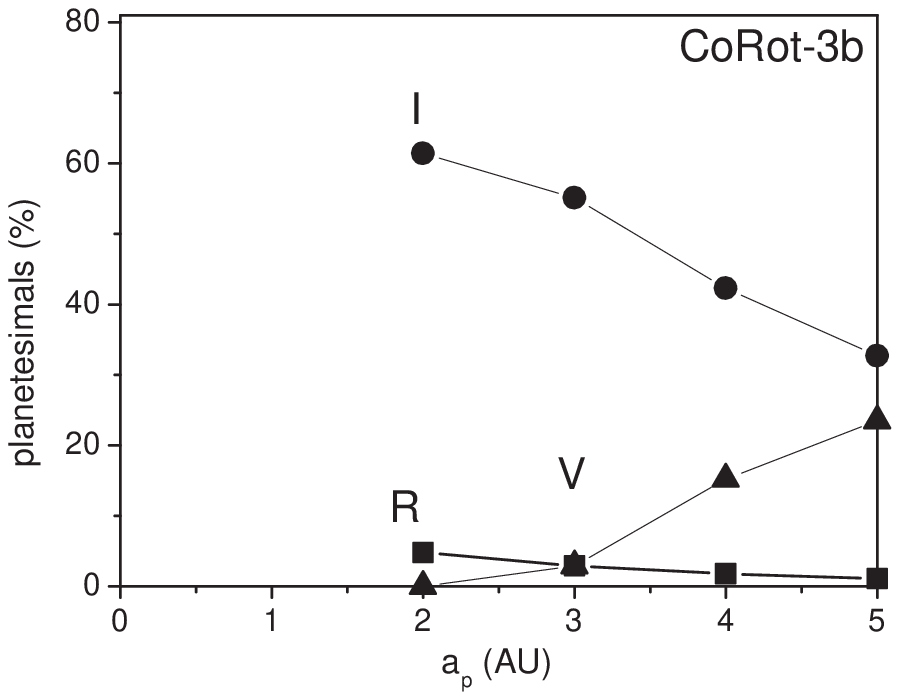}
\includegraphics[width=0.3\textwidth]{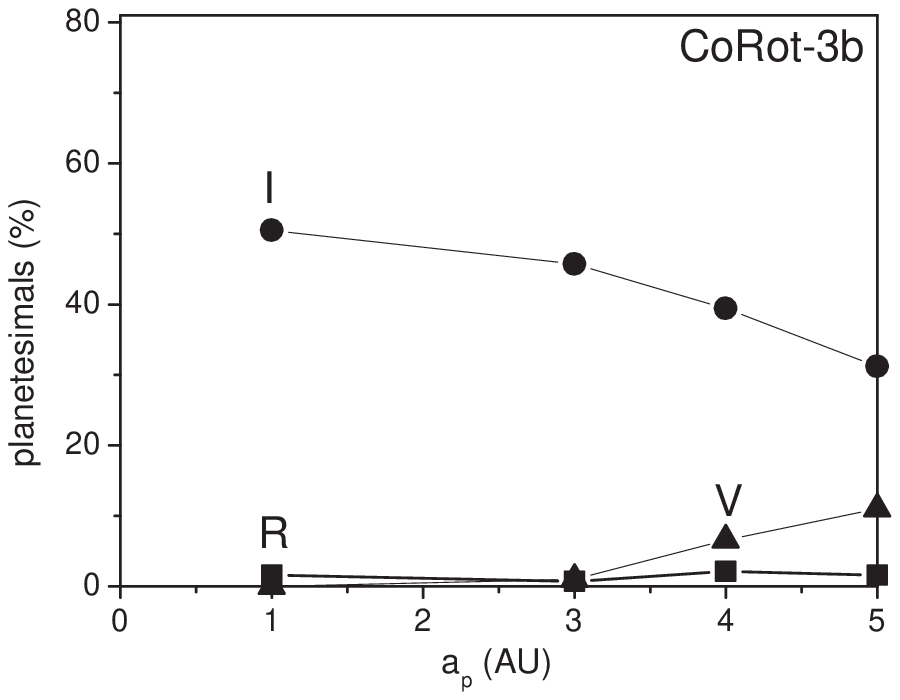}
\includegraphics[width=0.3\textwidth]{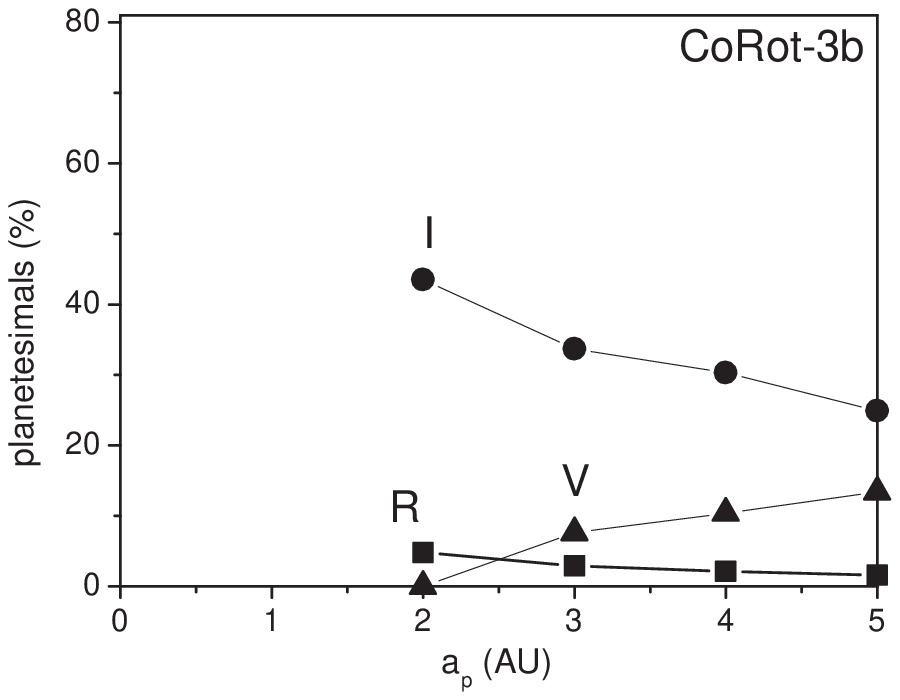}
\includegraphics[width=0.3\textwidth]{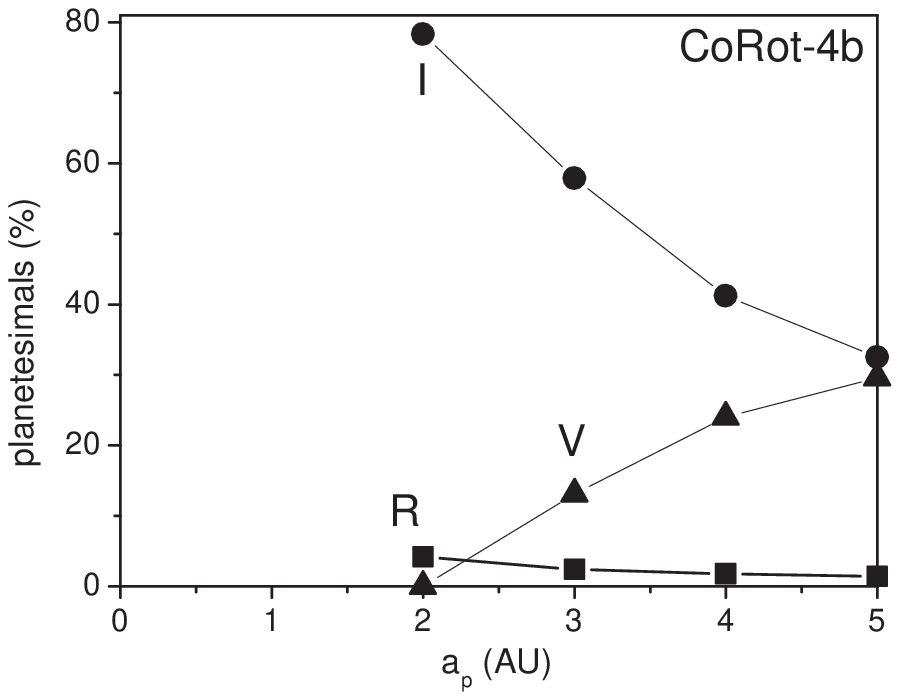}
\includegraphics[width=0.3\textwidth]{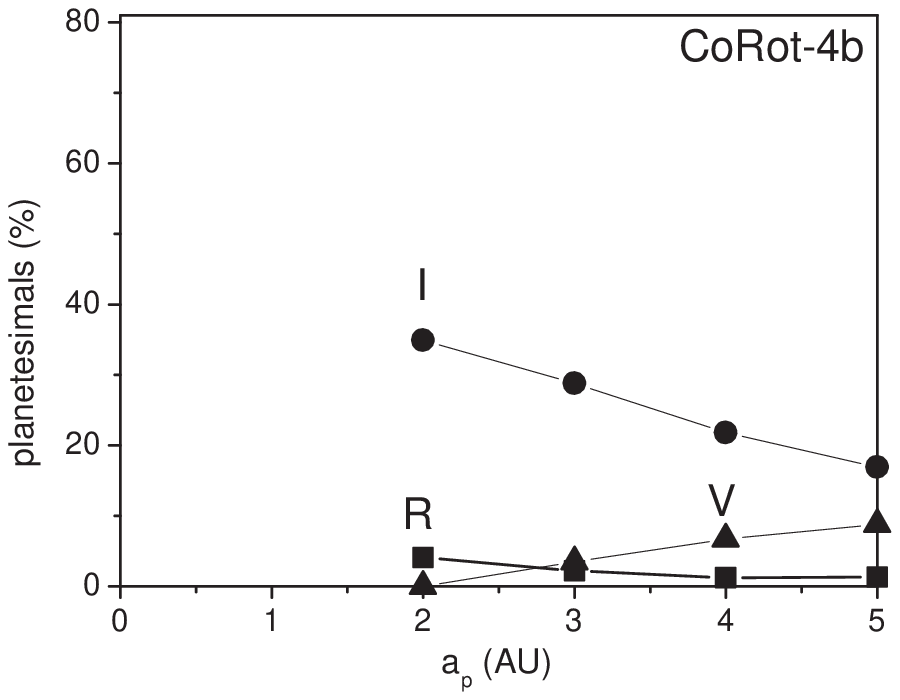}
\includegraphics[width=0.3\textwidth]{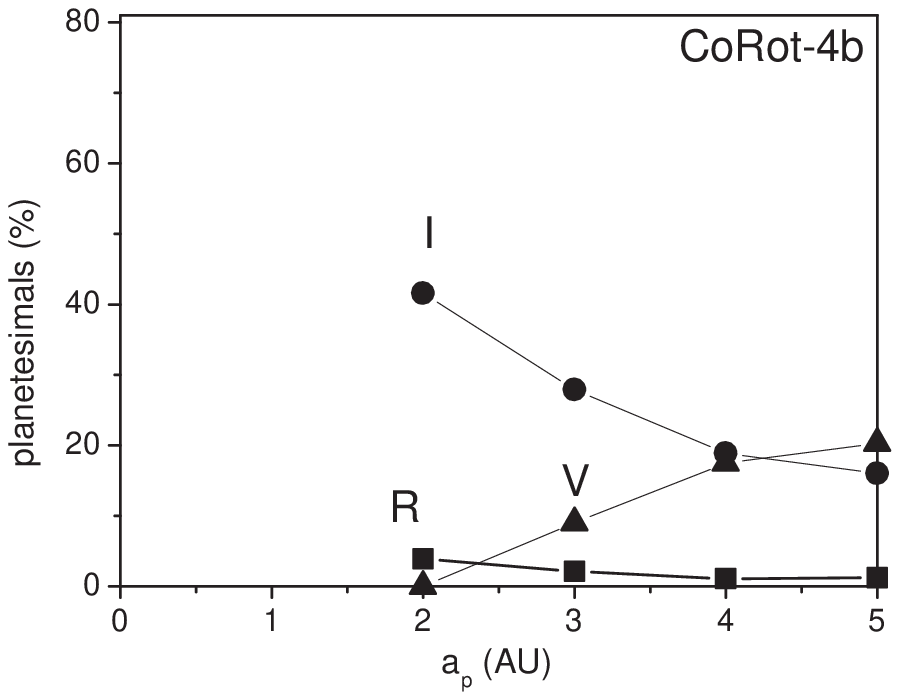}
 \caption{\label{f:5} Same as the first column of the Fig. \ref{f:4}, but  for planets initially at eccentric orbits.
 In the first, second and third  columns the results are for $e_{P}=0.1$, 0.3, and 0.5 cases, respectively.}
\end{center}
\end{figure*}

We also studied  the evolution of the system by considering the planet to be  on an eccentric orbit. 
In Fig. \ref{f:5}, we  present the results for $e_{P}=$ 0.1, 0.3, and 0.5, respectively. As expected 
for a circular orbit  (see first column of  Fig. \ref{f:4})  the percentage of 
collisions with the star   decreases (from about 10$\%$ to 35$\%$) when the planet's eccentricity is increased .
In general, we note that the percentage of intermediate and volatile materials accreted
decrease. For the CoRot-3 system, the integrations stopped when the last planetesimal was removed
from the system. The planet did not complete its migration at its current position. The same occurred for
the planet from CoRot-2 system when $e_{P}=$ 0.5. In both systems, the percentage values of refractory
material do not differ from those for $e_{P}=0$. In the  CoRot-4 case, some planetesimals had survived when
the integrations finished, but the planet arrived at its current position, $a_{D_{f}}=$ 0.09 AU. In this
case, we note an increase in the percentage of refractory material that fall on the star. Planetesimals in
the refractory zone are captured in mean motion resonances and some of them fall onto the star.

A planet on an eccentric orbit can significantly increase the eccentricities of the planetesimals because of
 its gravitational perturbation force. When planetesimals reach very high eccentricities they are then
removed from the system by collision or ejection. As shown in  \cite{winter07}, a fraction of planetesimals 
are  spread without being  captured in resonances. On the other hand, other planetesimals  are captured
 in several resonances (2:1, 3:1, 7:2, 5:1, 6:1, and 5:2). Planetesimals in higher resonances  are not so well
protected against close approaches with the planet and a fraction of them are ejected from the system.

\subsubsection{Warm planetesimals  disk}

\begin{figure*}[!]
\begin{center}
\includegraphics[width=0.3\textwidth]{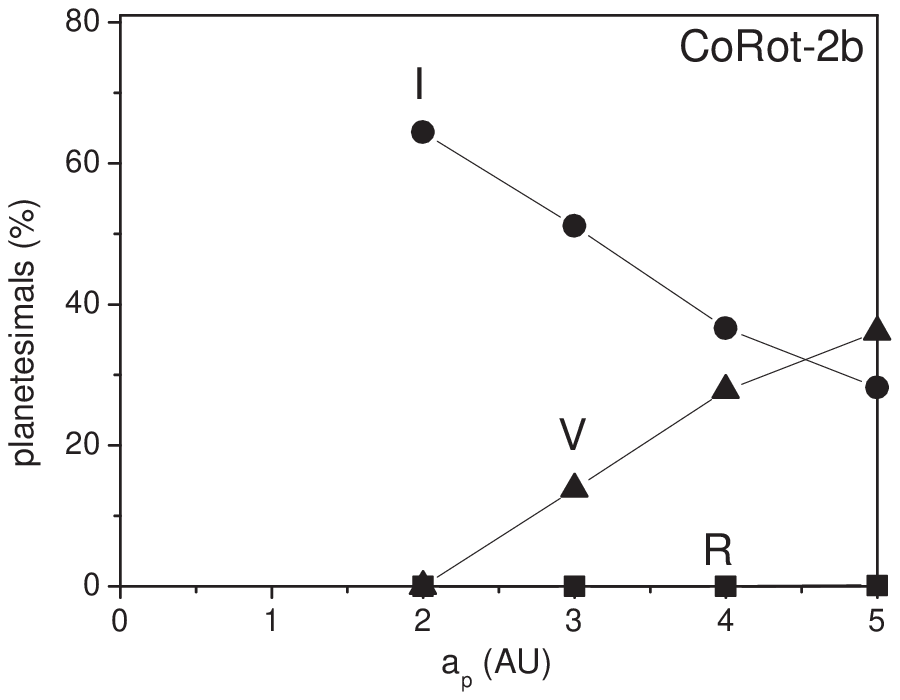}
\includegraphics[width=0.3\textwidth]{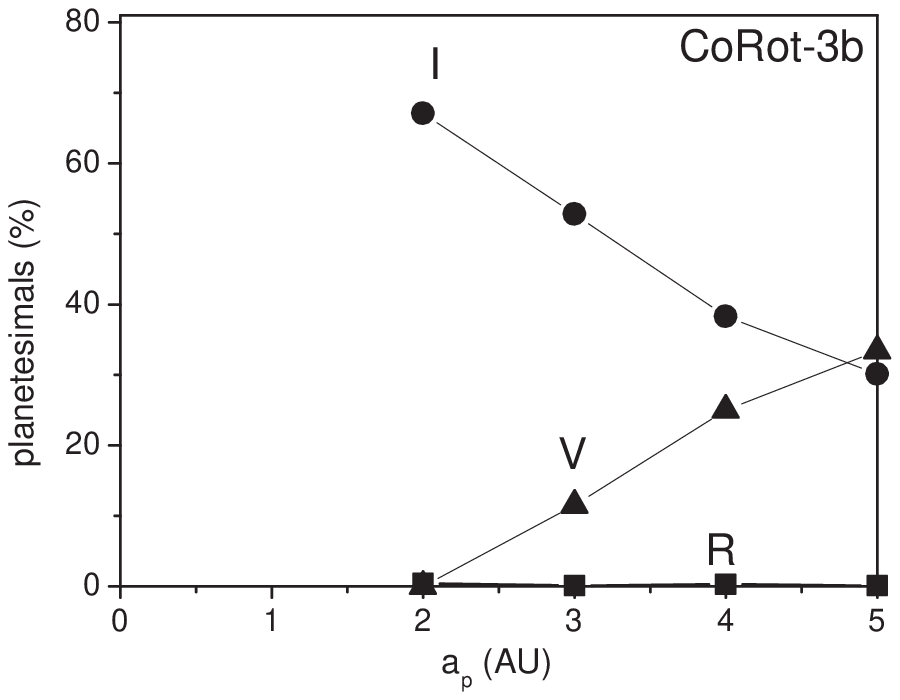}
\includegraphics[width=0.3\textwidth]{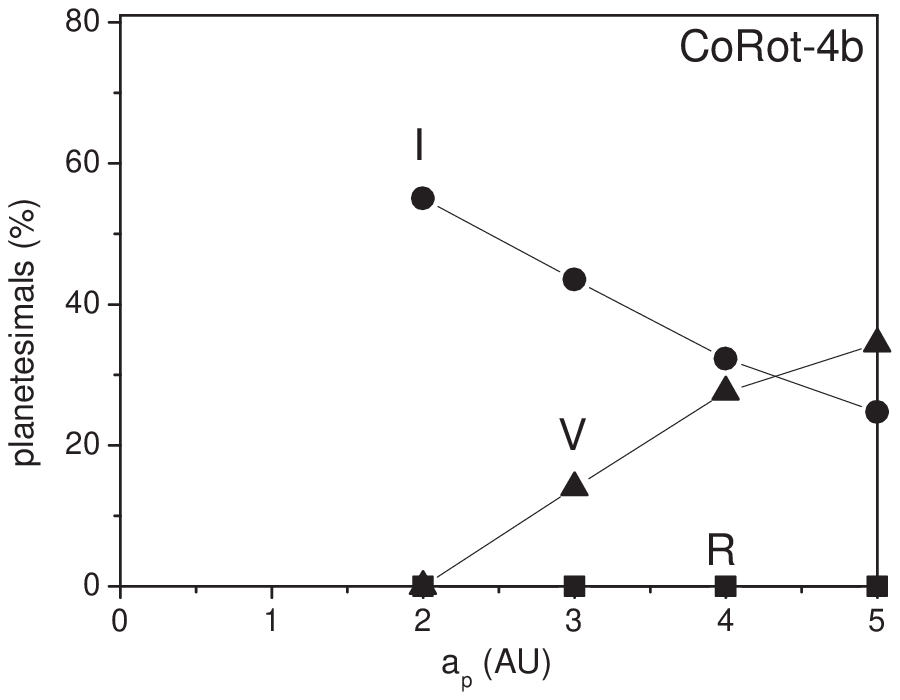}
 \caption{\label{f:6} Same as the first column of Fig. \ref{f:4}, but  for planetesimals initially distributed at 
eccentricities between 0 and 0.1.}
\end{center}
\end{figure*}

Here we consider a disk of  planetesimals with eccentricities  initially distributed between 0 and 0.1. The results 
are presented in Fig. \ref{f:6}. We note that  the results are similar to those of a disk of planetesimal initially in  
a circular orbit about a planet with an eccentricity of 0.1 (first column of Fig. \ref{f:5}).  The differences between these 
results and those from the first column of Fig. \ref{f:4} are due to the amount of refractory material that fall on the
 star. Here, only the CoRot-4 system has a percentage value of refractory material similar to those for $e_{P}=$0.0. In the
 present case, planetesimals in the refractory zone are ejected from the system. In the CoRot-2 and CoRot-3 systems, there
 is also a decrease in the amount of refractory material that falls onto the star, which was also ejected from the system. 
That is because the eccentricity (of the planetesimal or the planet) affects the capture in mean motion resonance
 between the planetesimal and the planet. We learned from  \cite{winter07} that the 2:1 resonance is responsible for the majority
 of  impacts onto the star. According to \cite{quillen06}, a body is captured into the 2:1 resonance when its initial eccentricity
 is below a certain limit given by $e_{lim}\sim 1.5\mu^{1/3}$. This is about 0.126, 0.220, and 0.371 for CoRoT-4, CoRoT-2, and 
CoRoT-3 systems, respectively.

\section{The link between migration and stellar abundances}

\begin{figure}                                                                                              
\resizebox{\hsize}{!}{\includegraphics{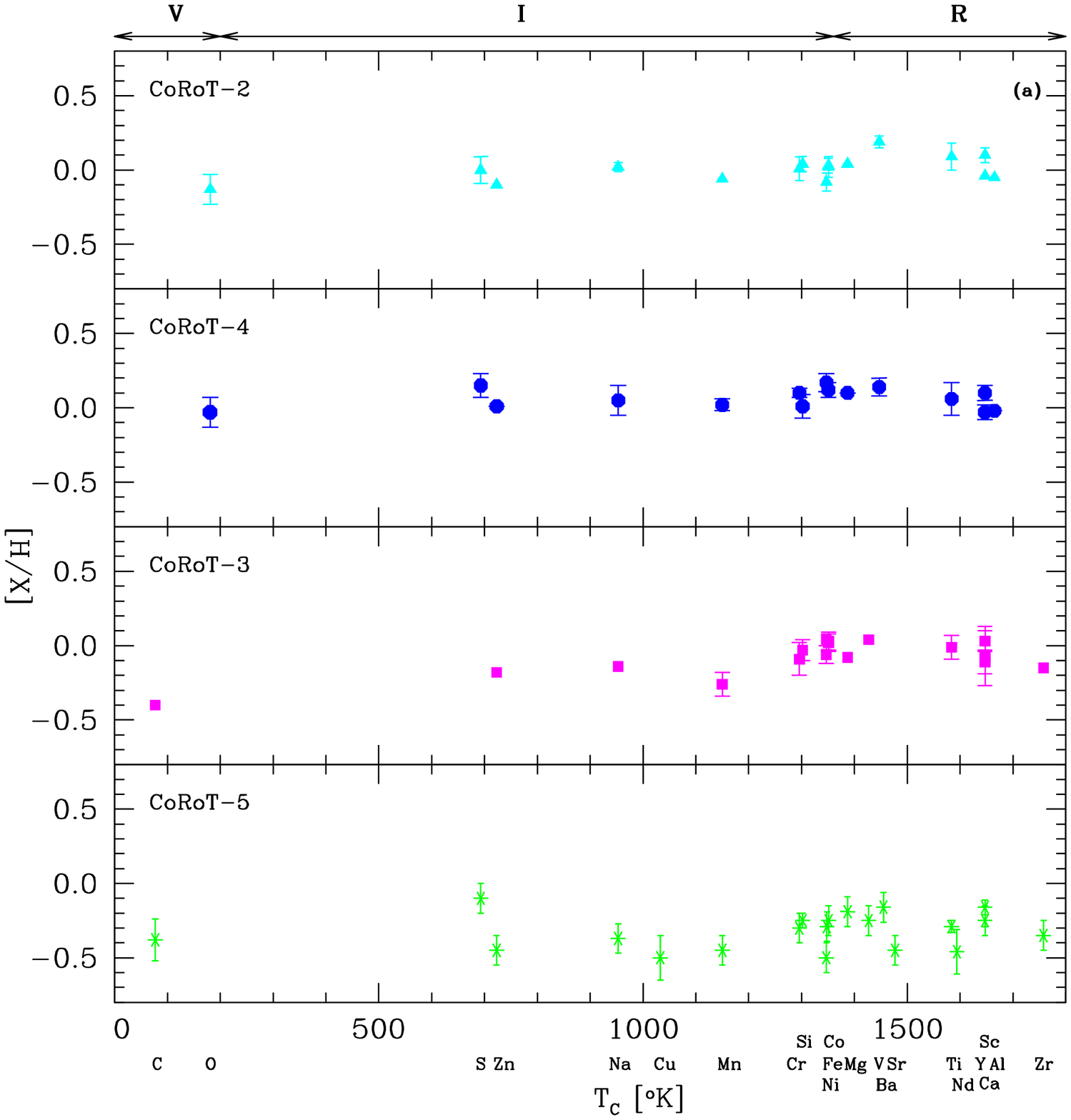}  }                                                   
\resizebox{\hsize}{!}{\includegraphics{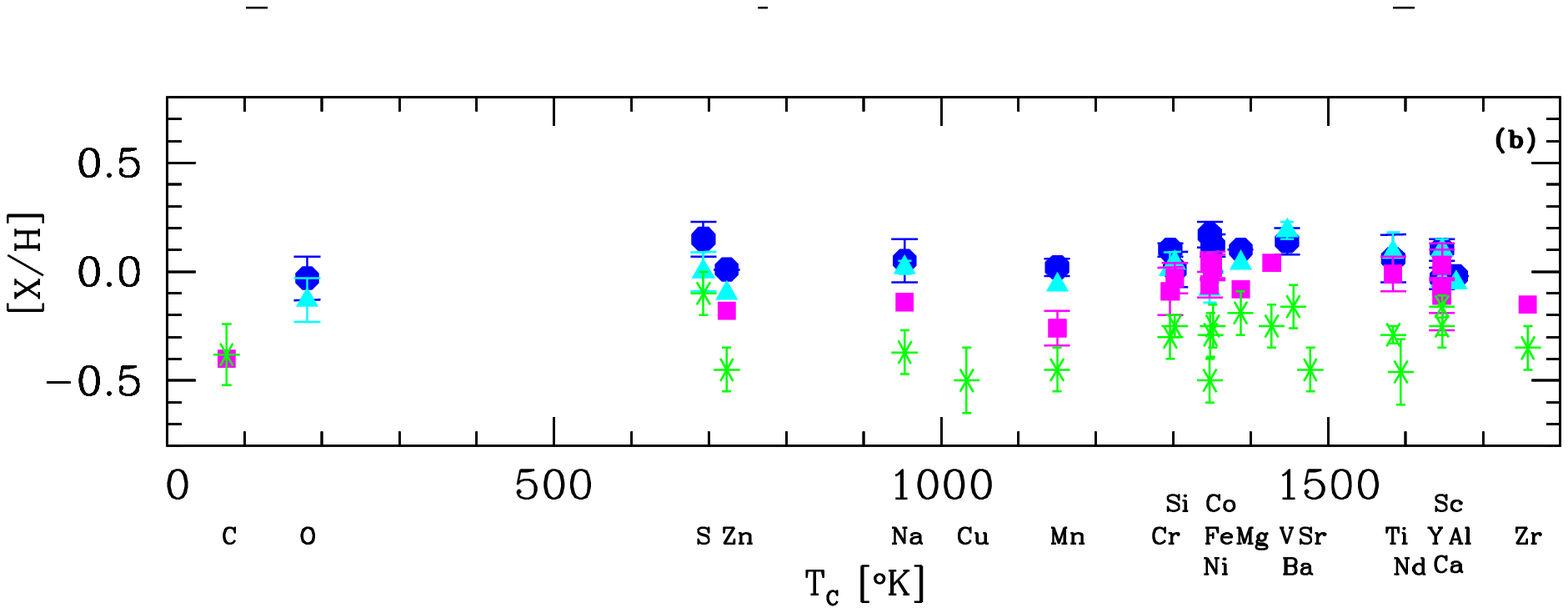} }
\vskip -165pt
                                                    
\caption{(a)- Relative abundances distribution of elements for the CoRoT stars as a function of their
  condensation temperatures (separated). (b)-  Superposed distributions: CoRoT-2 (cyan triangles), CoRoT-4
 (blue circles), CoRoT-3 (magenta squares) and CoRoT-5 (green stars).  }

 \label{f:temp}                                                                                                
\end{figure}

We now discuss some assumptions and physical conditions that are
the basis of the link between the consequences of the type of
planetary migration considered and the distributions of the
elemental abundances relative to the condensation temperatures $T_{\rm C}$.
To test this link,  we present  in Fig.  \ref{f:temp} the abundances 
 as a function of $T_{\rm C}$ for the CoRoT stars.

At a certain phase of the disk evolution, we assume that the disk is
homogeneously filled with planetesimals between a minimal typical
distance of $\sim$ 0.03 AU and a maximal one  of $\sim$ 4.5 AU. We also assume that a giant planet is
formed at $\sim$ 5 AU, before migration. An age of the system (star $+$
disk) of 20 to 30 Myr is critical for the efficiency of this process
\citep{winter07}. At this age, when the disk is already
devoid of its gas component, a low mass central star will indeed reduce, for the first time,
its external convective layer to its minimum configuration (see for instance \cite{ford99}).
In this situation, any metallic content of  the planetesimals falling onto the star will
probably remain in the photosphere of the star contributing to its metallic
enrichment. This may also be true for ages older than 30 Myr, but not younger than 20 Myr
because any enrichment would then be drastically reduced due to the larger
convective layer at these early epochs.

Clearly, the stellar enrichment depends on the quantity of infall
material. If this mass were of the order of 35 Earth masses 
of planetesimals consisting essentially of iron, this induced migration 
mechanism could furnish an iron enrichment of the order of 1.8 dex, as
observed in stars hosting hot Jupiters \citep{winter07}. But, how would this affect the
infall of a collection of rocky materials of varying  richnesses of refractory, intermediate, and volatile
elements? First, we must consider that in general, stars apparently without  detected
planets exhibit flat distributions (zero slope) of the element abundances with respect to $T_{\rm C}$  \citep{ecuvillon06}.
Then, in principle, any differential stellar accretion will produce a positive
slope of [X/H] versus $T_{\rm C}$ if the volatile contribution is smaller than
the refractory contribution. In contrast, negative slopes 
should appear if the volatile contribution is far more important.

Observationally speaking, also in the case of SWP, flat distributions
are universally found \citep{ecuvillon06}. It is mainly for this reason that zero gradient
distributions are, in the literature, synonymous with the absence of a
self-enrichment mechanism. Nevertheless, \cite{smith01}
found,  using a collection of abundance data from the literature, that
about six stars exhibit positive slopes. Other authors \citep{gonzalez06a,gonzalez06b,sadakane02,takeda01}
found  no dependence of the abundances on $T_{\rm C}$.

Applying our differential accretion mechanism methodology to three CoRoT systems (2, 3, and 4)
 considered in this paper, we found surprisingly  that in all cases the volatile and intermediate 
contributions were very similar and
much larger than the refractory ones (Fig. \ref{f:4}). Here, we
present the results for a relatively slow planetary migration of 10$^5$
years. As shown in \cite{winter07}, for shorter migration times,
the accretion mechanism is less efficient. We show this in
Fig. \ref{f:histo} only for the case of CoRoT-4, which involves in the infall of
intermediate material only.

At the top of Fig.  \ref{f:temp}, we have separated by arrows along the $T_{\rm C}$ scale, the three V, I, and R
domains. In general, we can infer that similar V and I contributions
represent a flat distribution of the elements in these two zones.
In case of  significant stellar accretion, the flat distribution will
only be shifted upward as a whole. In the case of modest or null accretion,
the original and flat distribution will be maintained. This is in
fact, the case of CoRoT systems 2 and 3, where migration, at least in
these conditions is not realized (see Sect. 6). Only in the case of 
CoRoT-4  may differentiate accretion can contribute approximately to a mild metal
enhancement,  but only for rapid migration (Fig. \ref{f:histo}).

We found that the accretion contribution  of highly refractory elements such as Ca, for example,
is small  compared to those of I and V elements. In this case, what would be
the meaning if, in case of a net metallic enhancement, we were to observe that the R elements have
the same levels of abundances as V
and I elements? We can indeed infer that R and I elements were
 originally  more mixed in the disk as we thought. In other words,
R elements are contained in I and maybe only  the two zones V and I are the more
realistic representatives ones.

We can interpret these results in another way. At the end of
Sect. 4.1, we discussed some effects of the mixing processes during the
primeval stages of a solar/stellar nebula. The detection of
silicate bands in comets \citep{hanner94} indicates that crystalline forms can exist
in these outer disk bodies \citep{bouwman00,wooden00}.
Since these crystals can only be formed at the relatively high disk
temperatures existing near the star, it has been proposed that a large-scale
 mixing occurred during the early disk epochs, transporting material
from the inner to the outer regions of the disk.
Mechanisms such as turbulence and large-scale meridional flows in regions close to
the disk midplane can act in this way. In contrast, inflow can exist above the midplane
\citep{keller04}. Whether these mechanisms can be changed by the
presence of magnetohydrodynamical turbulence was studied by
\cite{carballido05}. Independently,  we have ultimately identified for our case,
two different situations depending on the extent of
these mixing processes. On the  one hand, for a partial mixing we
maintain the R, I, and V zones or only V and I. 
On the other hand, we obtain a single large intermediate zone disk when the early disk is completely
and totally mixed.

\section{Results and conclusions}

In Fig. \ref{f:temp}, we present one of our most important result: 
the relative distribution of the abundances of refractory, intermediate,
and volatile elements as a function of their condensation temperatures
for four CoRoT stars. A flat distribution is found for all of
them, even if CoRoT-3 presents a slight tendency to display a positive
slope. Clearly, a more reliable determination of this slope would be
obtained if a complete collection of abundances of the highly volatile
CNO elements was at hand. In any case, we also note the importance to
dispose of measurements of the volatile S and Zn elements.

When considering the metallic abundances of CoRoT-2 and CoRoT-4,  derived 
in this work, we found that the metallicity of CoRoT-4 is
somewhat higher than that proposed by \cite{bouchy08}, hence confirming  a
mild metallic excess. We compare the stellar parameters
with those in the literature for CoRoT-2 and CoRoT-4  in Table  \ref{t:parameters}.

 CoRoT-2 star deserves  particular attention because of its significant stellar activity and its youth indicators
represented, respectively, by its Ca II H and K lines (Fig. \ref{f:ca}) and its Li resonance line 
(Fig. \ref{f:li-c2}), contrasting sharply with those of the older 1 Gyr star CoRoT-4 \citep{moutou08},
which does not exhibit  these features.
Its  age can be determined from the equivalent width (EW) and abundance of  Li
by using diagrams of the Li values as a function of
the $T_{\rm eff}$ values. By considering its measured EW of 124 m\AA\, we found, using
the diagrams presented by \cite{torres06,torres08}, that this star
falls in the lower dispersion limit of the Li abundances found for the members of the
Pleiades cluster, with an age of 119 $\pm$ 20 Myr \citep{ortega07}. By using its Li
abundance  found here of 2.6, we arrived at the same position in the diagrams of \cite{torres06}
 and \cite{dasilva09}. From all these matches, we infer an age
of 120 Myr for CoRoT-2, from which \cite{bouchy08} proposed an  age of $<$ 500 Myr.
This new age implies that of CoRoT-2 is  one of the youngest known two stars with planets, the
other being HD 70573, a debris-disk type star of a similar age and Li abundance
\citep{setiawan07}. The $T_{\rm eff}$ values of these two solar-type stars are in a short
temperature interval, where old solar-type stars with planets exhibit a peculiar Li
depletion with respect to similar stars without planets \citep{israelian09}. CoRoT-2
and HD 70573 are unaffected by this Li depletion due to their youth (see also \cite{pinsonneault09}.

We performed simulations of  planetary migration into an internal disk
composed of differentiated particles (planetesimals) with
refractory, intermediate, and volatile properties, and applied  these simulations to three
CoRoT systems. The accreted material contains large and similar
contributions of  {\it I} and  {\it V} particles and a very small contribution of
pure refractory elements, as shown in Fig. \ref{f:4}. In other
words, accretion is mainly ``cool'' and ``warm'' and not  ``hot'' as largely
mentioned in the literature. Because zone {\it I} exhibits a mixture of chemical properties,
mainly containing  common elements such as Fe, both intermediate and volatile
contributions produce a flat distribution of [X/H] versus $T_{\rm C}$,
as for the observed three CoRoT systems considered here.

When there is a complete primeval mixing of elements in the disk, a
single intermediate zone will be maintained  producing a flat distribution of
abundances as a function of $T_{\rm C}$. However, if the mixing is incomplete
or partial, some R and V rocky material will be maintained near
and far from the star, respectively. An extended I zone (see top of
Fig. \ref{f:temp}) will control a large part of the slope. In that case,
the agreement of the model with observations will  depend mainly on the
extended distribution of the element abundances  with $T_{\rm C}$ $<$ 304
K. Nevertheless, this region, containing volatile and highly
volatile elements such as CNO and noble gases, in general,  requires spectra of higher S/N
than our own to measure  the abundances.  By considering
our O abundances for CoRoT 2 and 4 and the C abundances for CoRoT 3 by
\cite{deleuil08} and that for CoRoT 5 \citep{rauer09}, we  obtained one collection
 of quite flat gradients   as presented in Fig. \ref{f:temp}. We note that
 CoRoT 3, for which the gradient is the least flat, is a peculiar case 
because  its ``planet'' is  probably a brown dwarf star. We conclude
that to obtain a reliable collection of highly volatile element
abundances of CNO and noble gases,  data of even higher S/N  are required.

In any case, independently of whether accretion is important or not, depending
on the accreting mass, the main result of this work, is that, for the CoRoT systems considered, flat
distributions that appear to be observationally the rule, do not
represent the absence of a self-enrichment mechanism as sometimes mentioned in the
literature. In contrast, we show here that flat distributions of the elements
abundances as function of  $T_{\rm C}$ could be a natural result of accretion.

Does the discussed type of migration represent true situations
for the  CoRoT systems 2, 3, and 4? This is not
the case for, at least systems 2 and 3, where unrealistically high disk
masses are necessary to bring the planets to their observed final
distances with respect to the star \citep{adams03}. System 4 can
be considered an exception. A different  migration of type II 
in a disk containing gas  \citep{papaloizou06,armitage07}
has to be invoked for CoRoT-2. The CoRoT-3 system is
particularly difficult and could represent a challenge to migration
theories. If this planetary body, with an exceptionally high mass of 21.66 M$_J$ is a real
planet and not a brown dwarf, a type II migration study must first
avoid the collision of the planet with the star due to the large
planetary eccentric orbits developed, which is the case, at least, for a
$\sim$ 10 M$_J$ \citep{rice08}. These authors mention that if the
gas disk dissipates quickly, the eccentric orbits of these massive planets
could eventually be  tidally circularized \citep{ford06}. Is this the case of
CoRoT-3 system?

CoRoT-4 system  with a planet of 0.72 M$_J$, also requires a high primeval disk mass of $\sim$  
0.2 -- 0.3 M$_{\odot}$. Even if it were an extreme case, it could be acceptable for a star of
1.2 M$_{\odot}$. Nevertheless, our simulations indicate that in this case only a very rapid migration of
1000 yr can produce a metal enhancement as observed of [I/H] $\sim$ 0.1., if occurred in the
first 20--30 Myr, when the stellar convective layer  for the first time attains its minimum configuration \citep{ford99}, 
This could also be possible in principle, even for the smaller planet of mass 0.467 M$_J$ related to CoRoT-5 \citep{rauer09}.
We also show that these results are highly dependent on the model adopted for the disk distribution 
regions in terms of refractory, intermediate, and volatile elements and the other parameters considered.

%::::::::::::::::::::::::::::::::::::::::::::::::::::::::::::::::::::::::::::

\begin{acknowledgements}
Part of this work was supported by CAPES, CNPq and FAPESP.
The authors would like to thank the referee for suggestions that improved the paper.
\end{acknowledgements}

%\bibliographystyle{aa}
%\bibliography{bilbioAbundances}

\end{document}